\documentclass[10pt,aps,pra,twocolumn,showpacs,showkeys,groupedaddress,amssymb,amsmath]{revtex4-1}

\pdfoutput=1

\usepackage{graphicx}  
\usepackage{dcolumn}   
\usepackage{bm}        

\usepackage{natbib}
\usepackage[colorlinks=true,urlcolor=blue,bookmarks=false]{hyperref}  

\usepackage{color}

\usepackage{verbatim}
\usepackage{amsmath}
\usepackage{latexsym}
\usepackage{amsfonts}
\usepackage{amssymb}

\begin{document}

\title{A Proposal for Experimental Detection of Amplitude $n$th-Power Squeezing}

\author{Ranjana Prakash}
\email{prakash\_ranjana1974@rediffmail.com}
\author{Ajay K. Yadav}
\email{ajaypdau@gmail.com}
\affiliation{Physics Department, University of Allahabad, Allahabad-211002, UP, India.}


\begin{abstract}
Recently, in several theoretical investigations, amplitude $n$th-power squeezing has been studied with $n = 2, 3, 4, 5$. In the present paper, we give a proposal for experimental detection of amplitude $n$th-power squeezing using ordinary homodyning with coherent light for arbitrary power $n$ and discuss in detail its theory. The proposed scheme requires only repeated measurements of the factorial moments of number of photons in the light obtained after homodyning, with various shifts of phase of coherent light, and involves no approximations, whatsoever. This has advantage over the method proposed by Shchukin and Vogel [Phys. Rev. A {\bf 72}, 043808 (2005)] in that our method requires only one beam splitter and only one photodetector, and also lesser number of repeatitions of experiment with phase-shifted coherent light.
\end{abstract}



\maketitle

\section{Introduction}

A single mode light has two components in quadrature and the operators corresponding to these in quantum mechanics are non-commuting  \cite{[{See, e.g., the review articles }] [] walls83306141, *teich891153,* dodonov0241, loudon8734709, leonhardt951989} and satisfy an uncertainty relation. For classical optical fields, which have a non-negative weight function in Sudarshan-Glauber diagonal representation \cite{sudarshan6310277, *glauber631312766}, the variances of the quadrature operators have equal lower bounds. Optical fields may have a non-classical feature and variance of one quadrature amplitude may be less than this lower bound at the expense of increased variance for the other. This non-classical feature, called squeezing, was studied earlier in academic interest \cite{mollow671601076, *mollow671601097, *chandra7041196, *raiford7021541, *chandra719409, *chandra719677, *chandra719688, *chandra719767, *prakash7492167, *prakash75757, *prakash75763, *prakash761441, *prakash761448} but its importance has now been understood because of its application to optical communication \cite{yuen7824657}, optical waveguide tap \cite{shapiro805351}, gravitational wave detection \cite{caves81231693, *pace93473173, *mckenzie0493161105}, interferometric techniques \cite{caves85313068, *schumaker85313093}, enhancing the channel capacity \cite{saleh87582656}, quantum teleportation \cite{braunstein9880869, *zhang0062064302, *bowen0350801}, quantum dense coding \cite{ban9919, *braunstein0061042302}, quantum cryptography \cite{hillery0061022309}, Nano-displacement measurement \cite{treps046664}, optical storage \cite{hsu06421001} and amplification of signals \cite{kozlovskii073774, *laflamme1183033803}. Generation of squeezed states has been reported in a variety of nonlinear optical processes, e.g., multi-photon absorption \cite{simaan758539, *loudon844967}, degenerate parametric amplifier \cite{milburn8139401}, free electron laser \cite{becker8248475}, harmonic generation \cite{mandel8232437, *lugiato838256}, degenerate parametric oscillation \cite{milburn8327392}, degenerate four-wave mixing \cite{bondurant8430343, *reid85311622}, degenerate hyper-Raman scattering \cite{perinova8450401}, resonance fluorescence \cite{ficek84292004}, the single atom-single field mode interaction \cite{li0119797, *zheng07273460}, superposed coherent states \cite{prakash03319305, *prakash04341201, *prakash111221058}, and nonlinear beam splitter \cite{prakash10352212}.

Squeezing is understood by writing the annihilation operator $\hat{a}$ in terms of the two hermitian quadrature amplitude operators $\hat{X}_{1}$ and $\hat{X}_{2}$ in the form, $\hat{a} = \hat{X}_{1} + i\hat{X}_{2}$ or $\hat{X}_{1} = \frac{1}{2} (\hat{a}^{\dagger} + \hat{a})$, $\hat{X}_{2} = \frac{i}{2} (\hat{a}^{\dagger} - \hat{a})$ which gives $[\hat{X}_{1}, \hat{X}_{2}] = \frac{i}{2}$ or $\langle(\Delta \hat{X}_{1})^{2}\rangle\langle(\Delta \hat{X}_{2})^{2}\rangle \geq \frac{1}{16}$, where conical brackets denote expectation values and $\Delta\hat{X}_{1,2} \equiv \hat{X}_{1,2} - \langle\hat{X}_{1,2}\rangle$. $\hat{X}_{1}$ is said to be squeezed if $\langle(\Delta \hat{X}_{1})^{2}\rangle < \frac{1}{4}$ and $\hat{X}_{2}$ is said to be squeezed if $\langle(\Delta \hat{X}_{2})^{2}\rangle < \frac{1}{4}$.

Instead of considering $\hat{X}_{1}$ and $\hat{X}_{2}$ one can consider the most general quadrature amplitude operator, $\hat{X}_{\theta} = \hat{X}_{1} \cos\theta + \hat{X}_{2} \sin\theta = \frac{1}{2} (\hat{a}^{\dagger} e^{i\theta} + \hat{a} e^{-i\theta})$. The commutation relation, $[\hat{X}_{\theta}, \hat{X}_{\theta + \pi/2}] = \frac{i}{2}$, which gives uncertainty relation, $\langle(\Delta \hat{X}_{\theta})^{2}\rangle\langle(\Delta \hat{X}_{\theta + \pi/2})^{2}\rangle \geq \frac{1}{16}$, then leads to idea of squeezing of the general component $\hat{X}_{\theta}$ when $\langle(\Delta\hat{X}_{\theta})^{2}\rangle < \frac{1}{4}$. The non-classical nature of squeezing, is clear from the fact that, for Sudarshan-Glauber diagonal representation \cite{sudarshan6310277, glauber631312766}, $\rho = \int d^{2}\alpha P(\alpha) |\alpha\rangle\langle\alpha|$, we have
\begin{eqnarray}
\langle(\Delta\hat{X}_{\theta})^{2}\rangle - \frac{1}{4} &=& \int d^{2}\alpha P(\alpha) [Re\{(\alpha - \bar{\alpha})e^{-i\theta}\}]^{2} < 0, \nonumber\\
\bar{\alpha} &\equiv& \int d^{2}\alpha P(\alpha) \alpha.
\end{eqnarray}

Concepts of amplitude squeezing and phase squeezing were introduced using (i) the uncertainty relation $\langle(\Delta\hat{N})^{2}\rangle\langle(\Delta\hat{\phi})^{2}\rangle \geq \frac{1}{4}$, where photon number operator and phase uncertainty were defined by $\hat{N} = \hat{a}^{\dagger}\hat{a}$, $\langle(\Delta\hat{\phi})^{2}\rangle = [\langle(\Delta\hat{S})^{2}\rangle + \langle(\Delta\hat{C})^{2}\rangle]/[\langle\hat{S}\rangle^{2} + \langle\hat{C}\rangle^{2}]$ with $\hat{C} + i\hat{S} = (\hat{N} + 1)^{-1/2} \hat{a}$, \cite{carruthers6514387} and (ii) the relation $\langle(\Delta\hat{N})^{2}\rangle \geq \langle\hat{N}\rangle$ for classical light, resulting in the definition $\langle(\Delta\hat{N})^{2}\rangle < \langle\hat{N}\rangle$ for amplitude squeezing and $\langle(\Delta\hat{\phi})^{2}\rangle < 1/4\langle\hat{N}\rangle$ for phase squeezing. The terminology, amplitude and phase squeezing, is widely accepted and is described in text-books \cite{mandelandwolf95, *scullyandzubairy97, *gerryandknight05, *wallsandmilburn08} and encyclopedia \cite{[] [{ available at \url{http://www.rp-photonics.com/amplitude_squeezed_light.html}; }]asl, *[] [{ available at \url{http://www.rp-photonics.com/squeezed_states_of_light.html}.}]ssl}.

Squeezing of radiation has been generalized in a number of ways. In the first, Hong and Mandel \cite{hong8554323, *hong8532974} consider even powers of difference between quadrature amplitudes and its mean values, and the field is called squeezed to order $2n$ whenever $\langle(\Delta\hat{X}_{\theta})^{2n}\rangle$ is less than its value for any coherent state. This type of squeezing has been studied by several authors \cite{fernandez86118400, *tombesi88374778, *marian91443325, *marian92452044, *wang00392583, *praksah03342769, *mishra102833284, *prakash11284289}. Another generalization was introduced by Hillery \cite{hillery8762135, *hillery87363796} for the lowest order ($n = 2$), and by Zhang et al. \cite{zhang9015027} for $n > 2$. These authors consider generalized quadrature operators obtained by separating Hermitian and anti-Hermitian parts of the square or $n$th-power of the annihilation operator and define squeezing.

Explicitly, in the general case, we define
\begin{equation}
\hat{X}^{(n)}_{\theta} = \frac{1}{2} (\hat{a}^{\dagger n} e^{i\theta} + \hat{a}^{n} e^{-i\theta}),
\end{equation}
which gives commutation relation
\begin{equation}
[\hat{X}^{(n)}_{\theta}, \hat{X}^{(n)}_{\theta + \pi/2}] = \frac{i}{2} \hat{W}^{(n)},
\end{equation}
where
\begin{equation}
\hat{W}^{(n)} = \sum^{n}_{r=1} r! ({^{n}C_{r}})^{2} \hat{a}^{\dagger n-r}\hat{a}^{n-r}.
\end{equation}
The uncertainty relation,
\begin{equation}
\langle(\Delta\hat{X}^{(n)}_{\theta})^{2}\rangle\langle(\Delta\hat{X}^{(n)}_{\theta + \pi/2})^{2}\rangle \geq \frac{1}{16} \langle\hat{W}^{(n)}\rangle^{2},
\end{equation}
therefore,  leads to squeezing of $\hat{X}^{(n)}_{\theta}$ when
\begin{equation}
\langle(\Delta\hat{X}^{(n)}_{\theta})^{2}\rangle < \frac{1}{4} \langle\hat{W}^{(n)}\rangle.
\label{canps1}
\end{equation}

Since we can write $\langle:(\hat{X}^{(n)}_{\theta})^{2}:\rangle = \langle(\hat{X}^{(n)}_{\theta})^{2}\rangle - \frac{1}{4} \langle\hat{W}^{(n)}\rangle$, we have $\langle:(\Delta\hat{X}^{(n)}_{\theta})^{2}:\rangle = \langle(\Delta\hat{X}^{(n)}_{\theta})^{2}\rangle - \frac{1}{4} \langle\hat{W}^{(n)}\rangle$, where double dots (: :) denote normal ordering of operators. This tells that Eq.~(\ref{canps1}), which gives the definition of $n$th-order squeezing, can also be written in the simpler form,
\begin{equation}
\langle:(\Delta\hat{X}^{(n)}_{\theta})^{2}:\rangle < 0.
\label{canps2}
\end{equation}

This type of squeezing was called amplitude-squared squeezing by Hillery \cite{hillery8762135, *hillery87363796} and $n$th-order squeezing by Zhang et al. \cite{zhang9015027}. The spirit in this generalization of squeezing is entirely different from that of lowest order amplitude squeezing, as in the former the uncertainty relations for generalized quadrature operators are used while for the latter number-phase uncertainty relations are used. Just after the publication of papers of Hillery \cite{hillery8762135, *hillery87363796} and Zhang et al. \cite{zhang9015027}, two papers of Zhan \cite{zhan91160498, *zhan9246686} used the phrase ``Amplitude $N$th-Power Squeezing'' for such generalization and since then this phrase has been used invariably in all later publications \cite{gerry88371779, mahran89404476, *obada929199, *prakash213621, giri04691, *giri0822219, prakash06201458, *wu0747933, marian97553051, *prakash0846359, prakash0538665, *mishra07112859, kumar9681053, *giri06202265, *rani0739157, *sen08551697, du93482198, *wang957917, *rani0739735, *rani08281341, *rani09232681, giri05191943, *rani07277427, wang954247}.

Amplitude-squared squeezing has been studied for an anharmonic oscillator \cite{gerry88371779}, the interaction between atom and radiation field \cite{mahran89404476, *obada929199, *prakash213621}, mixing of waves \cite{giri04691, *giri0822219}, Kerr medium \cite{prakash06201458, *wu0747933}, coherent states \cite{ marian97553051, *prakash0846359}. Also enhancement of amplitude-squared squeezing is studied by mixing it with coherent light beam \cite{prakash0538665, *mishra07112859}. Amplitude $n$th-power squeezing for higher-order has been studied by several authors with $n = 3$ \cite{kumar9681053, *giri06202265, *rani0739157, *sen08551697}, 4 \cite{du93482198, *wang957917, *rani0739735, *rani08281341, *rani09232681}, 5 \cite{giri05191943, *rani07277427}, and for $n = k$ \cite{wang954247}.

It may be noted that since the Hong and Mandel's concept of $2n$th order squeezing is not based on any uncertainty relation, $2n$th order squeezing can be obtained for both $\hat{X}_{\theta}$ and $\hat{X}_{\theta + \pi/2}$ simultaneously \cite{lynch86334431, *hong86334432, *lynch94492800}. Simultaneous squeezing of $\hat{X}^{(n)}_{\theta}$ and $\hat{X}^{(n)}_{\theta + \pi/2}$ is however ruled out because of the uncertainty relations. Hillery also considered a second type of amplitude squared squeezing by separating $(\hat{a} - \langle\hat{a}\rangle)^{2}$ into its Hermitian and anti-Hermitian parts. A different type of amplitude $n$th-power squeezing has been defined by Bužek and Jex \cite{buzek90414079}. Other generalizations of squeezing involve multi-mode operators and their commutation relations; the common examples being sum and difference squeezing \cite{hillery89403147, *kumar97136441, *kumar9810485, *prakash0745363}, spin squeezing \cite{prakash057757}, atomic squeezing \cite{prakash0742475}, and polarization squeezing \cite{chirkin9323870, *psicop2009, *prakash112843568}.

Proposal for experimental detection of ordinary squeezing was given by Mandel \cite{[] [ (see also reference \cite{mandelandwolf95, *scullyandzubairy97, *gerryandknight05, *wallsandmilburn08} for homodyning).] mandel8249136} using homodyning with intense light with adjustable phase from a local oscillator and measuring the number of photons and its square in one of the outputs. Such homodyning of squeezed light converts squeezing into sub-Poissonian photon statistics \cite{mandel794205, *short8351384, *kumar1088181} and the degree of squeezing is obtained from measurements of expectation values of photon number operator and its square. Prakash and Kumar \cite{prakash057786} showed a similar conversion of fourth-order squeezing into second-order sub-Poissonian photon statistics and proposed a balanced homodyne method for detection of fourth-order squeezed light in the similar fashion. Prakash and Mishra \cite{prakash06392291, *prakash07402531} extended the proposal of ordinary homodyning for experimental detection of amplitude-squared squeezing by measuring higher order moments of number operator of mixed light with shifted phases. They also studied higher-order sub-Poissonian photon statistics conditions for non-classicality and discussed its use for the detection of Hong and Mandel’s squeezing of arbitrary order. Prakash et al. reported recently \cite{prakash10245547} an ordinary homodyne method for detection of second type of amplitude-squared squeezing of Hillery by measuring the higher-order moments of the number operator in light obtained by homodyning with intense coherent light. Another proposal for the detection of amplitude $k$th-power squeezing (\cite{shchukin0572043808} for k = 2, \cite{vogel0784012020} for general) by the measurements of the moments $\langle\hat{a}^{k}\rangle,\langle\hat{a}^{2k}\rangle$ and $\langle\hat{a}^{\dagger k}\hat{a}^{k}\rangle$ has been given on a technique based on balanced homodyne correlation measurement \cite{shchukin0572043808, vogel0784012020, shchukin0696200403}. In this method number of beam splitters and photodetectors required increase with the increase in order of moments. Also a very large number of repeated measurements with phase shift $\varphi$ of the local oscillator are required as evaluation of Fourier transform of a function of $\varphi$ is involved. There is no experimental demonstration of any form of higher-order squeezing in the literature. But proposals for experimental detection of some higher-order squeezing have been given \cite{prakash057786, prakash06392291, prakash07402531, prakash10245547, shchukin0572043808, vogel0784012020}.

We present here a proposal for experimental detection of amplitude $n$th-power squeezing for any arbitrary power $n$ by using the ordinary homodyne detection method. It may be noted that we work without taking any approximation whatsoever like replacing operator for coherent state by $c$-number or taking transmittance or reflectance of the beam splitter very small, and our method requires only one photodetector and only one beam splitter.

\section{The detection scheme}

\begin{figure}
	\centering
		\includegraphics[width=0.48\textwidth]{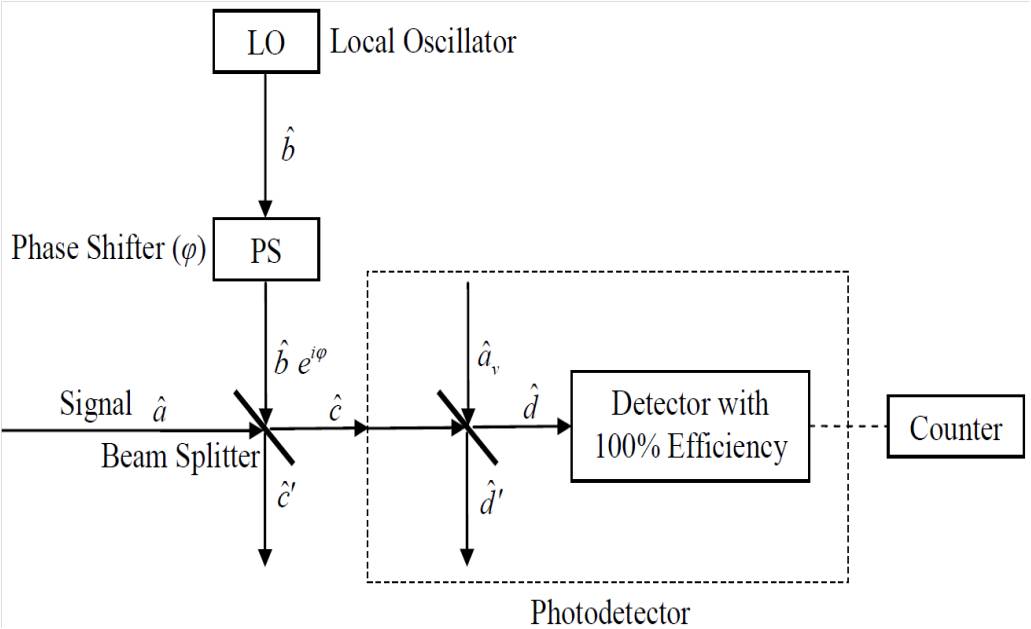}
	\caption{Schematic diagram of detection of squeezing via ordinary homodyning.}
	\label{oh}
\end{figure}

Schematic diagram for proposed detection scheme, shown in Fig.~\ref{oh}, contains the experimental outline and also the conceptual meaning of quantum efficiency of the experimental detector \cite{[{See, e.g., the review article }] [] loudon8734709}. The beam splitter and ideal detector placed inside dotted rectangle model the real photodetector \cite{loudon8734709}. The signal represented by operator $\hat{a}$ is mixed using beam splitter with signal $\hat{b}e^{i\varphi}$ obtained by shifting by $\varphi$ the phase of signal from a local oscillator represented by operator $\hat{b}$ to give output signal $\hat{c}$ and $\hat{c}'$ One of the output signals, say $\hat{c}$ is detected. If the beam splitter has transmittance $T$ and we write $t=\sqrt{T}$ and  as $r=\sqrt{1-T}$ coefficients of transmission and reflection for the amplitudes, respectively, we can write \cite{[{See, e.g., the review article }] [] leonhardt951989}
\begin{equation}
\hat{c}=t\hat{a}+r\hat{b}e^{i\varphi},
\end{equation}
with $t$ and $r$ real. Number operator of the mixed light is then
\begin{equation}
\hat{N}_{c}=\hat{c}^{\dagger}\hat{c}=(t\hat{a}^{\dagger}+r\hat{b}^{\dagger}e^{-i\varphi})(t\hat{a}+r\hat{b}e^{i\varphi}).
\end{equation}
In this model \cite{loudon8734709} of photodetector with quantum efficiency $\eta$, the beam splitter mixes (i) input signal $\hat{c}$ and (ii) vacuum signal $\hat{a}_{v}$ and one of the outputs $\hat{d}$ given by
\begin{equation}
\hat{d}=\sqrt{\eta}\hat{c}+\sqrt{1-\eta}\hat{a}_{v}
\label{outputd}
\end{equation}
is detected by an ideal detector having $100\%$ efficiency. It is interesting to note that this model explains that the detected counts, $\langle \hat{d}^{\dagger}\hat{d} \rangle$, are $\eta$ times the incident number of photons, $\langle \hat{c}^{\dagger}\hat{c} \rangle$, i.e.,
\begin{equation}
\langle \hat{d}^{\dagger}\hat{d} \rangle=\eta\langle \hat{c}^{\dagger}\hat{c} \rangle
\end{equation}
which is also obtained directly from Eq.~(\ref{outputd}). For factorial moments of order $n$, we get similarly
\begin{equation}
\langle \hat{d}^{\dagger n}\hat{d}^{n} \rangle=\eta^{n}\langle \hat{c}^{\dagger n}\hat{c}^{n} \rangle.
\end{equation}
The relationship for $n$th-power of photon number operator is not a simple one. This is one reason why we involve factorial moments of photon number operator and not powers of the photon number operator in our analysis and result, as was done earlier \cite{prakash06392291, *prakash07402531}.

For the setup under consideration, the observed factorial moments of counts with phase shift $\varphi$ is then
\begin{eqnarray}
M^{(n)}_{\varphi} &=& \eta^{n}\langle\hat{c}^{\dagger n}\hat{c}^{n}\rangle \nonumber \\
&=& \eta^{n}\sum^{n}_{l,m=0}{^{n}C_{l}}{^{n}C_{m}}t^{2n-l-m}r^{l+m}\nonumber \\
&& \langle \hat{a}^{\dagger n-l}\hat{a}^{n-m}\hat{b}^{\dagger l}\hat{b}^{m} \rangle e^{i(m-l)\varphi}.
\end{eqnarray}
If the local oscillator gives output in the coherent state $|\beta\rangle$ with the complex amplitude $\beta=|\beta|e^{i\theta_{\beta}}$ and observations are done for $M^{(n)}_{\varphi}$ for $\varphi = k\pi/n$ with $k = 0, 1,..., 2n-1$ we can easily find the values of observables
\begin{eqnarray}
P_{n} &\equiv& \frac{1}{2n} \sum^{2n-1}_{k = 0} M^{(n)}_{k\pi/n} e^{ik \pi} \nonumber \\
&=& (\eta t r |\beta|)^{n} [\langle \hat{a}^{\dagger n} \rangle e^{in \theta_{\beta}} + \langle \hat{a}^{n} \rangle e^{-in \theta_{\beta}}], 
\label{pn}
\end{eqnarray}
\begin{eqnarray}
Q_{n} &\equiv& \frac{1}{2n} \sum^{2n-1}_{k = 0} M^{(n)}_{k\pi/n} \nonumber \\
&=& \eta^{n} t^{2n} \sum^{n}_{l=0} ({^{n}C_{l}})^{2} (t^{-1} r |\beta|)^{2l} \langle \hat{a}^{\dagger n - l}\hat{a}^{n - l} \rangle. 
\label{qn}
\end{eqnarray}
Eq.~(\ref{pn}) gives a method for measuring $\langle \hat{X}^{(n)}_{\theta} \rangle$ using $\theta_{\beta} = \theta / n$, but a method for measuring  $\langle \hat{W}^{(n)} \rangle$ and therefore for moments $\langle \hat{a}^{\dagger n - l}\hat{a}^{n - l} \rangle$ is still to be desired. To achieve this end, we can solve Eq.~(\ref{qn}) for factorial moments of photon number operator by iteration. This is done in Appendix and it leads to
\begin{equation}
\langle \hat{a}^{\dagger n}\hat{a}^{n} \rangle = \eta^{-n} t^{-2n} \sum^{n}_{s = 0} K_{s} (^{n}C_{s})^{2} \eta^{s} (r |\beta|)^{2s} Q_{n-s},
\label{fmno}
\end{equation}
where $K_{s}$ is defined by
\begin{equation}
K_{s} = - \sum^{s - 1}_{l = 0} K_{l} ({^{s}C_{l}})^{2} \text{ with } K_{0} = 1.
\label{ks}
\end{equation}
Straight forward calculations lead to
\begin{eqnarray}
&& P_{2n} - P^{2}_{n} + 2(\eta t r |\beta|)^{2n} \langle \hat{a}^{\dagger n}\hat{a}^{n} \rangle \nonumber \\
&& = P_{2n} - P^{2}_{n} + 2 \sum^{n}_{s = 0} K_{s} (^{n}C_{s})^{2} \eta^{n + s} (r |\beta|)^{2(n + s)} Q_{n-s} \nonumber \\
&& = 4(\eta t r |\beta|)^{2n} [\langle : (\Delta \hat{X}^{(n)}_{\theta})^{2} : \rangle \cos^{2}(n \theta_{\beta} - \theta) \nonumber \\
&& + \langle : (\Delta \hat{X}^{(n)}_{\theta + \pi / 2})^{2} : \rangle \sin^{2}(n \theta_{\beta} - \theta) + \frac{1}{2} (\langle \Delta \hat{X}^{(n)}_{\theta} \Delta \hat{X}^{(n)}_{\theta + \pi / 2} \rangle \nonumber \\
&& + \langle \Delta \hat{X}^{(n)}_{\theta + \pi / 2} \Delta \hat{X}^{(n)}_{\theta} \rangle) \sin 2(n \theta_{\beta} - \theta)] \nonumber \\
&& = 
\begin{cases}
4(\eta t r |\beta|)^{2n} \langle : (\Delta \hat{X}^{(n)}_{\theta})^{2} : \rangle, \text{ if } (n \theta_{\beta} - \theta) = 0; \\
4(\eta t r |\beta|)^{2n} \langle : (\Delta \hat{X}^{(n)}_{\theta + \pi / 2})^{2}: \rangle, \text{ if } (n \theta_{\beta} - \theta) = \frac{\pi}{2}.
\end{cases}
\label{anps}
\end{eqnarray}
This equation shows that amplitude $n$th-power squeezing for any value of $\theta$, defined by Eqs. (6) or (7), can be detected by measurement of observables $P_{n}, Q_{n}, \eta, T$ and $|\beta|$.

\section{Discussion of results}

We explained how $P_{n}$ and $Q_{n}$ can be found from measurements of $\langle \hat{N}^{(n)}_{\varphi} \rangle$ (see Eqs.~(\ref{pn}) and ~(\ref{qn}). Measurement of quantum efficiency $\eta$ for a given detector is somewhat tricky and requires use of spontaneous parametric down conversion \cite{penin91303582, *castelletto9532501, *brida9835397, *migdall02412914, *lu0213186}. If a photon of a large energy $\hbar\omega$ breaks to create two photons of energies $\hbar\omega_{1}$ and $\hbar\omega_{2}$ (with $\omega_{1} + \omega_{2} = \omega$), the latter two photons should give coincidence counts only ideally. If \textit{N} photons break and the quantum efficiencies for modes $\omega_{1}$ and $\omega_{2}$ are $\eta_{1}$ and $\eta_{2}$ (both $< 1$), the experiment registers counts $N_{1} = \eta_{1} N, N_{2} = \eta_{2} N$ and coincidence counts $N_{c} = \eta_{1} \eta_{2} N$. The quantum efficiencies are then $\eta_{1} = N_{c}/N_{2}$ and $\eta_{2} = N_{c}/N_{1}$.

Once $\eta$ is determined for any detector, the detector can be used to measure $T$ and $|\beta|$ easily.

It should be noted that as $n$ increases the choice of $\theta_{\beta}$ becomes more sensitive and for detection of $\langle:(\Delta\hat{X}^{(n)}_{\theta})^{2}:\rangle$, $\theta_{\beta}$ is to be set as $\theta/n$.

For $n = 1$, i.e., for ordinary squeezing detection was first studied by Mandel \cite{mandel8249136} (see also references \cite{mandelandwolf95, *scullyandzubairy97, *gerryandknight05, *wallsandmilburn08} for homodyning). His formula is different from that obtained from Eq.~(\ref{anps}) after substituting $n = 1$. This is understandable as Mandel's result is approximate, and is obtained under the approximation, $|\beta| >> 1/r >> 1$, in our notations. Also, for $n = 2$, i.e., for amplitude-squared squeezing, the results of Prakash and Mishra \cite{prakash06392291, *prakash07402531} are different from the results obtained from Eq.(\ref{anps}) as their results are also approximate, the same approximation ($|\beta| >> 1/r >> 1$, in our notations) being used.

Another method for detecting amplitude $k$th-power squeezing (\cite{vogel0784012020}; see also \cite{shchukin0572043808}) has been proposed by Shchukin and Vogel based on measurement of some correlation functions from which the values of $\langle\hat{a}^{k}\rangle,\langle\hat{a}^{2k}\rangle$ and $\langle\hat{a}^{\dagger k}\hat{a}^{k}\rangle$ can be inferred. The authors arranged several beam splitters in $d$ levels ($r$th level having $2^{r-1}$ beam splitters and $d$ being called the depth) between an entrance beam splitter and a sequence of $2^{d}$ photodetectors. Since an additional beam splitter is used before the entrance beam splitter, total  $2^{d}+1$ beam splitters are required. If $k$ satisfies $2^{n} \geq k > 2^{n-1}$, for obtaining values of $\langle\hat{a}^{k}\rangle$ and $\langle\hat{a}^{\dagger k}\hat{a}^{k}\rangle$ depth $d = n$ is needed, which requires use of $2^{n} + 1$ beam splitters and $2^{n}$ photodetectors. For obtaining the values of $\langle\hat{a}^{2k}\rangle$ obviously, depth $n+1$ and, therefore, $2^{n+1}+1$ beam splitters and $2^{n+1}$ photodetectors are required. In the method proposed in the present paper, only one beam splitter and only one photodetector is required. The Shchukin-Vogel method proposes measurement of correlations with several phase shifts $\varphi$ of local oscillator. Since a Fourier transform of correlations over $\varphi$ is required for inferring the values of moments $\langle\hat{a}^{k}\rangle$ and $\langle\hat{a}^{\dagger k}\hat{a}^{k}\rangle$, very large number of repeatition of experiments with changed values of $\varphi$ should be required. It may be noted that in the method proposed in the present paper only $4k$ repeatitions are required.

\begin{acknowledgments}
We would like to thank Prof. H. Prakash and Prof. N. Chandra for their interest and critical comments. Also we would like to acknowledge Dr. Rakesh Kumar, Dr. Pankaj Kumar, Dr. Devendra K. Mishra, Dr. Namrata Shukla, Dr. Manoj K. Mishra, Mr. Ajay K. Maurya and Mr. Vikram Verma for their valuable and stimulating discussions. One of the authors (AKY) is grateful to the University Grants Commission(UGC), New Delhi, India for financial support.
\end{acknowledgments}

\appendix*\section{}

We can separate the $\langle\hat{a}^{\dagger n}\hat{a}^{n}\rangle$ term on right hand side of Eq.~(\ref{qn}) and write
\begin{eqnarray}
\langle\hat{a}^{\dagger n}\hat{a}^{n}\rangle &=& \eta^{-n} t^{-2n} Q_{n} - \sum^{n}_{l=1} K_{0} ({^{l}C_{0}})^{2} ({^{n}C_{l}})^{2} \nonumber \\
&& (t^{-1} r |\beta|)^{2l} \langle\hat{a}^{\dagger n-l}\hat{a}^{n-l}\rangle
\label{A1}
\end{eqnarray}
with $K_{0} = 1$. In the first term in summation on the right hand side we substitute for $\langle\hat{a}^{\dagger n-1}\hat{a}^{n-1}\rangle$ the expression obtained from Eq.~(\ref{A1}) using $K_{0} ({^{l}C_{0}})^{2} = 1$ and this gives
\begin{eqnarray*}
&& \langle\hat{a}^{\dagger n}\hat{a}^{n}\rangle = \eta^{-n} t^{-2n} Q_{n} \nonumber \\
&& + K_{1} ({^{n}C_{1}})^{2} (t^{-1} r |\beta|)^{2} [\eta^{-n+1} t^{-2n+2} Q_{n-1} \nonumber \\
&& - \sum^{n-1}_{m=1}({^{n-1}C_{m}})^{2} (t^{-1} r |\beta|)^{2m} \langle\hat{a}^{\dagger n-m-1}\hat{a}^{n-m-1}\rangle] \nonumber \\
&& - \sum^{n}_{l=2} K_{0} ({^{l}C_{0}})^{2} ({^{n}C_{l}})^{2} (t^{-1} r |\beta|)^{2l} \langle \hat{a}^{\dagger n-l}\hat{a}^{n-l} \rangle
\end{eqnarray*}
with $K_{1} = - K_{0} ({^{l}C_{0}})^{2}$. This can be simplified and written as
\begin{eqnarray}
\langle\hat{a}^{\dagger n}\hat{a}^{n}\rangle &=& \eta^{-n} t^{-2n} [Q_{n} + K_{1} ({^{n}C_{1}})^{2} \eta (r |\beta|)^{2} Q_{n-1}] \nonumber \\
&& - \sum^{n}_{l=2} [K_{0} ({^{l}C_{0}})^{2} + K_{1} ({^{l}C_{1}})^{2}] \nonumber \\
&& ({^{n}C_{l}})^{2} (t^{-1} r |\beta|)^{2l} \langle \hat{a}^{\dagger n-l}\hat{a}^{n-l} \rangle.
\end{eqnarray}
If we again substitute the first term in summation the expression obtained for $\langle\hat{a}^{\dagger n-2}\hat{a}^{n-2}\rangle$ from Eq.~(\ref{A1}) and simplify, we get
\begin{eqnarray}
&& \langle\hat{a}^{\dagger n}\hat{a}^{n}\rangle = \eta^{-n} t^{-2n} [Q_{n} + K_{1} ({^{n}C_{1}})^{2} \eta (r |\beta|)^{2} Q_{n-1} \nonumber \\
&& + K_{2} ({^{n}C_{2}})^{2} \eta^{2} (r |\beta|)^{4} Q_{n-2}] - \sum^{n}_{l=3} [K_{0} ({^{l}C_{0}})^{2} + K_{1} ({^{l}C_{1}})^{2} \nonumber \\
&& + K_{2} ({^{l}C_{2}})^{2}] ({^{n}C_{l}})^{2} (t^{-1} r |\beta|)^{2l} \langle \hat{a}^{\dagger n-l}\hat{a}^{n-l} \rangle.
\end{eqnarray}
with $K_{2} = - [K_{0} ({^{l}C_{0}})^{2} + K_{1} ({^{l}C_{1}})^{2}]$. If we go on doing similar exercises we get required Eq.~(\ref{fmno}), where $K_{s}$ is defined by Eq.~(\ref{ks}).

\bibliography{references_03}

\begin{thebibliography}{130}%
\makeatletter
\providecommand \@ifxundefined [1]{%
 \@ifx{#1\undefined}
}%
\providecommand \@ifnum [1]{%
 \ifnum #1\expandafter \@firstoftwo
 \else \expandafter \@secondoftwo
 \fi
}%
\providecommand \@ifx [1]{%
 \ifx #1\expandafter \@firstoftwo
 \else \expandafter \@secondoftwo
 \fi
}%
\providecommand \natexlab [1]{#1}%
\providecommand \enquote  [1]{``#1''}%
\providecommand \bibnamefont  [1]{#1}%
\providecommand \bibfnamefont [1]{#1}%
\providecommand \citenamefont [1]{#1}%
\providecommand \href@noop [0]{\@secondoftwo}%
\providecommand \href [0]{\begingroup \@sanitize@url \@href}%
\providecommand \@href[1]{\@@startlink{#1}\@@href}%
\providecommand \@@href[1]{\endgroup#1\@@endlink}%
\providecommand \@sanitize@url [0]{\catcode `\\12\catcode `\$12\catcode
  `\&12\catcode `\#12\catcode `\^12\catcode `\_12\catcode `\%12\relax}%
\providecommand \@@startlink[1]{}%
\providecommand \@@endlink[0]{}%
\providecommand \url  [0]{\begingroup\@sanitize@url \@url }%
\providecommand \@url [1]{\endgroup\@href {#1}{\urlprefix }}%
\providecommand \urlprefix  [0]{URL }%
\providecommand \Eprint [0]{\href }%
\providecommand \doibase [0]{http://dx.doi.org/}%
\providecommand \selectlanguage [0]{\@gobble}%
\providecommand \bibinfo  [0]{\@secondoftwo}%
\providecommand \bibfield  [0]{\@secondoftwo}%
\providecommand \translation [1]{[#1]}%
\providecommand \BibitemOpen [0]{}%
\providecommand \bibitemStop [0]{}%
\providecommand \bibitemNoStop [0]{.\EOS\space}%
\providecommand \EOS [0]{\spacefactor3000\relax}%
\providecommand \BibitemShut  [1]{\csname bibitem#1\endcsname}%
\let\auto@bib@innerbib\@empty
\bibitem [{\citenamefont {Walls}(1983)}]{walls83306141}%
  \BibitemOpen
  \bibfield  {author} {\bibinfo {author} {\bibfnamefont {D.~F.}\ \bibnamefont
  {Walls}},\ }\href@noop {} {\bibfield  {journal} {\bibinfo  {journal}
  {Nature}\ }\textbf {\bibinfo {volume} {306}},\ \bibinfo {pages} {141}
  (\bibinfo {year} {1983})}\BibitemShut {NoStop}%
\bibitem [{\citenamefont {Teich}\ and\ \citenamefont
  {Saleh}(1989)}]{teich891153}%
  \BibitemOpen
  \bibfield  {author} {\bibinfo {author} {\bibfnamefont {M.~C.}\ \bibnamefont
  {Teich}}\ and\ \bibinfo {author} {\bibfnamefont {B.~E.~A.}\ \bibnamefont
  {Saleh}},\ }\href@noop {} {\bibfield  {journal} {\bibinfo  {journal} {Quantum
  Opt.}\ }\textbf {\bibinfo {volume} {1}},\ \bibinfo {pages} {153} (\bibinfo
  {year} {1989})}\BibitemShut {NoStop}%
\bibitem [{\citenamefont {Dodonov}(2002)}]{dodonov0241}%
  \BibitemOpen
  \bibfield  {author} {\bibinfo {author} {\bibfnamefont {V.~V.}\ \bibnamefont
  {Dodonov}},\ }\href@noop {} {\bibfield  {journal} {\bibinfo  {journal} {J.
  Opt. B: Quantum Semiclass. Opt.}\ }\textbf {\bibinfo {volume} {4}},\ \bibinfo
  {pages} {R1} (\bibinfo {year} {2002})}\BibitemShut {NoStop}%
\bibitem [{\citenamefont {Loudon}\ and\ \citenamefont
  {Knight}(1987)}]{loudon8734709}%
  \BibitemOpen
  \bibfield  {author} {\bibinfo {author} {\bibfnamefont {R.}~\bibnamefont
  {Loudon}}\ and\ \bibinfo {author} {\bibfnamefont {P.~L.}\ \bibnamefont
  {Knight}},\ }\href@noop {} {\bibfield  {journal} {\bibinfo  {journal} {J.
  Mod. Opt.}\ }\textbf {\bibinfo {volume} {34}},\ \bibinfo {pages} {709}
  (\bibinfo {year} {1987})}\BibitemShut {NoStop}%
\bibitem [{\citenamefont {Leonhardt}\ and\ \citenamefont
  {Paul}(1995)}]{leonhardt951989}%
  \BibitemOpen
  \bibfield  {author} {\bibinfo {author} {\bibfnamefont {U.}~\bibnamefont
  {Leonhardt}}\ and\ \bibinfo {author} {\bibfnamefont {H.}~\bibnamefont
  {Paul}},\ }\href@noop {} {\bibfield  {journal} {\bibinfo  {journal} {Prog.
  Qunat. Electr.}\ }\textbf {\bibinfo {volume} {19}},\ \bibinfo {pages} {89}
  (\bibinfo {year} {1995})}\BibitemShut {NoStop}%
\bibitem [{\citenamefont {Sudarshan}(1963)}]{sudarshan6310277}%
  \BibitemOpen
  \bibfield  {author} {\bibinfo {author} {\bibfnamefont {E.~C.~G.}\
  \bibnamefont {Sudarshan}},\ }\href@noop {} {\bibfield  {journal} {\bibinfo
  {journal} {Phys. Rev. Lett.}\ }\textbf {\bibinfo {volume} {10}},\ \bibinfo
  {pages} {277} (\bibinfo {year} {1963})}\BibitemShut {NoStop}%
\bibitem [{\citenamefont {Glauber}(1963)}]{glauber631312766}%
  \BibitemOpen
  \bibfield  {author} {\bibinfo {author} {\bibfnamefont {R.~J.}\ \bibnamefont
  {Glauber}},\ }\href@noop {} {\bibfield  {journal} {\bibinfo  {journal} {Phys.
  Rev.}\ }\textbf {\bibinfo {volume} {131}},\ \bibinfo {pages} {2766} (\bibinfo
  {year} {1963})}\BibitemShut {NoStop}%
\bibitem [{\citenamefont {Mollow}\ and\ \citenamefont
  {Glauber}(1967{\natexlab{a}})}]{mollow671601076}%
  \BibitemOpen
  \bibfield  {author} {\bibinfo {author} {\bibfnamefont {B.~R.}\ \bibnamefont
  {Mollow}}\ and\ \bibinfo {author} {\bibfnamefont {R.~J.}\ \bibnamefont
  {Glauber}},\ }\href@noop {} {\bibfield  {journal} {\bibinfo  {journal} {Phys.
  Rev.}\ }\textbf {\bibinfo {volume} {160}},\ \bibinfo {pages} {1076} (\bibinfo
  {year} {1967}{\natexlab{a}})}\BibitemShut {NoStop}%
\bibitem [{\citenamefont {Mollow}\ and\ \citenamefont
  {Glauber}(1967{\natexlab{b}})}]{mollow671601097}%
  \BibitemOpen
  \bibfield  {author} {\bibinfo {author} {\bibfnamefont {B.~R.}\ \bibnamefont
  {Mollow}}\ and\ \bibinfo {author} {\bibfnamefont {R.~J.}\ \bibnamefont
  {Glauber}},\ }\href@noop {} {\bibfield  {journal} {\bibinfo  {journal} {Phys.
  Rev.}\ }\textbf {\bibinfo {volume} {160}},\ \bibinfo {pages} {1097} (\bibinfo
  {year} {1967}{\natexlab{b}})}\BibitemShut {NoStop}%
\bibitem [{\citenamefont {Chandra}\ and\ \citenamefont
  {Prakash}(1970)}]{chandra7041196}%
  \BibitemOpen
  \bibfield  {author} {\bibinfo {author} {\bibfnamefont {N.}~\bibnamefont
  {Chandra}}\ and\ \bibinfo {author} {\bibfnamefont {H.}~\bibnamefont
  {Prakash}},\ }\href@noop {} {\bibfield  {journal} {\bibinfo  {journal} {Lett.
  Nuovo. Cim.}\ }\textbf {\bibinfo {volume} {4}},\ \bibinfo {pages} {1196}
  (\bibinfo {year} {1970})}\BibitemShut {NoStop}%
\bibitem [{\citenamefont {Raiford}(1970)}]{raiford7021541}%
  \BibitemOpen
  \bibfield  {author} {\bibinfo {author} {\bibfnamefont {M.~T.}\ \bibnamefont
  {Raiford}},\ }\href@noop {} {\bibfield  {journal} {\bibinfo  {journal} {Phys.
  Rev. A}\ }\textbf {\bibinfo {volume} {2}},\ \bibinfo {pages} {1541} (\bibinfo
  {year} {1970})}\BibitemShut {NoStop}%
\bibitem [{\citenamefont {Chandra}\ and\ \citenamefont
  {Prakash}(1971{\natexlab{a}})}]{chandra719409}%
  \BibitemOpen
  \bibfield  {author} {\bibinfo {author} {\bibfnamefont {N.}~\bibnamefont
  {Chandra}}\ and\ \bibinfo {author} {\bibfnamefont {H.}~\bibnamefont
  {Prakash}},\ }\href@noop {} {\bibfield  {journal} {\bibinfo  {journal}
  {Indian J. Pure Appl. Phys.}\ }\textbf {\bibinfo {volume} {9}},\ \bibinfo
  {pages} {409} (\bibinfo {year} {1971}{\natexlab{a}})}\BibitemShut {NoStop}%
\bibitem [{\citenamefont {Chandra}\ and\ \citenamefont
  {Prakash}(1971{\natexlab{b}})}]{chandra719677}%
  \BibitemOpen
  \bibfield  {author} {\bibinfo {author} {\bibfnamefont {N.}~\bibnamefont
  {Chandra}}\ and\ \bibinfo {author} {\bibfnamefont {H.}~\bibnamefont
  {Prakash}},\ }\href@noop {} {\bibfield  {journal} {\bibinfo  {journal}
  {Indian J. Pure Appl. Phys.}\ }\textbf {\bibinfo {volume} {9}},\ \bibinfo
  {pages} {677} (\bibinfo {year} {1971}{\natexlab{b}})}\BibitemShut {NoStop}%
\bibitem [{\citenamefont {Chandra}\ and\ \citenamefont
  {Prakash}(1971{\natexlab{c}})}]{chandra719688}%
  \BibitemOpen
  \bibfield  {author} {\bibinfo {author} {\bibfnamefont {N.}~\bibnamefont
  {Chandra}}\ and\ \bibinfo {author} {\bibfnamefont {H.}~\bibnamefont
  {Prakash}},\ }\href@noop {} {\bibfield  {journal} {\bibinfo  {journal}
  {Indian J. Pure Appl. Phys.}\ }\textbf {\bibinfo {volume} {9}},\ \bibinfo
  {pages} {688} (\bibinfo {year} {1971}{\natexlab{c}})}\BibitemShut {NoStop}%
\bibitem [{\citenamefont {Chandra}\ and\ \citenamefont
  {Prakash}(1971{\natexlab{d}})}]{chandra719767}%
  \BibitemOpen
  \bibfield  {author} {\bibinfo {author} {\bibfnamefont {N.}~\bibnamefont
  {Chandra}}\ and\ \bibinfo {author} {\bibfnamefont {H.}~\bibnamefont
  {Prakash}},\ }\href@noop {} {\bibfield  {journal} {\bibinfo  {journal}
  {Indian J. Pure Appl. Phys.}\ }\textbf {\bibinfo {volume} {9}},\ \bibinfo
  {pages} {767} (\bibinfo {year} {1971}{\natexlab{d}})}\BibitemShut {NoStop}%
\bibitem [{\citenamefont {Prakash}\ \emph {et~al.}(1974)\citenamefont
  {Prakash}, \citenamefont {Chandra},\ and\ \citenamefont
  {Vachaspati}}]{prakash7492167}%
  \BibitemOpen
  \bibfield  {author} {\bibinfo {author} {\bibfnamefont {H.}~\bibnamefont
  {Prakash}}, \bibinfo {author} {\bibfnamefont {N.}~\bibnamefont {Chandra}}, \
  and\ \bibinfo {author} {\bibnamefont {Vachaspati}},\ }\href@noop {}
  {\bibfield  {journal} {\bibinfo  {journal} {Phys. Rev. A}\ }\textbf {\bibinfo
  {volume} {9}},\ \bibinfo {pages} {2167} (\bibinfo {year} {1974})}\BibitemShut
  {NoStop}%
\bibitem [{\citenamefont {Prakash}\ \emph
  {et~al.}(1975{\natexlab{a}})\citenamefont {Prakash}, \citenamefont
  {Chandra},\ and\ \citenamefont {Vachaspati}}]{prakash75757}%
  \BibitemOpen
  \bibfield  {author} {\bibinfo {author} {\bibfnamefont {H.}~\bibnamefont
  {Prakash}}, \bibinfo {author} {\bibfnamefont {N.}~\bibnamefont {Chandra}}, \
  and\ \bibinfo {author} {\bibnamefont {Vachaspati}},\ }\href@noop {}
  {\bibfield  {journal} {\bibinfo  {journal} {Indian J. Pure Appl. Phys.}\
  }\textbf {\bibinfo {volume} {13}},\ \bibinfo {pages} {757} (\bibinfo {year}
  {1975}{\natexlab{a}})}\BibitemShut {NoStop}%
\bibitem [{\citenamefont {Prakash}\ \emph
  {et~al.}(1975{\natexlab{b}})\citenamefont {Prakash}, \citenamefont
  {Chandra},\ and\ \citenamefont {Vachaspati}}]{prakash75763}%
  \BibitemOpen
  \bibfield  {author} {\bibinfo {author} {\bibfnamefont {H.}~\bibnamefont
  {Prakash}}, \bibinfo {author} {\bibfnamefont {N.}~\bibnamefont {Chandra}}, \
  and\ \bibinfo {author} {\bibnamefont {Vachaspati}},\ }\href@noop {}
  {\bibfield  {journal} {\bibinfo  {journal} {Indian J. Pure Appl. Phys.}\
  }\textbf {\bibinfo {volume} {13}},\ \bibinfo {pages} {763} (\bibinfo {year}
  {1975}{\natexlab{b}})}\BibitemShut {NoStop}%
\bibitem [{\citenamefont {Prakash}\ \emph
  {et~al.}(1976{\natexlab{a}})\citenamefont {Prakash}, \citenamefont
  {Chandra},\ and\ \citenamefont {Vachaspati}}]{prakash761441}%
  \BibitemOpen
  \bibfield  {author} {\bibinfo {author} {\bibfnamefont {H.}~\bibnamefont
  {Prakash}}, \bibinfo {author} {\bibfnamefont {N.}~\bibnamefont {Chandra}}, \
  and\ \bibinfo {author} {\bibnamefont {Vachaspati}},\ }\href@noop {}
  {\bibfield  {journal} {\bibinfo  {journal} {Indian J. Pure Appl. Phys.}\
  }\textbf {\bibinfo {volume} {14}},\ \bibinfo {pages} {41} (\bibinfo {year}
  {1976}{\natexlab{a}})}\BibitemShut {NoStop}%
\bibitem [{\citenamefont {Prakash}\ \emph
  {et~al.}(1976{\natexlab{b}})\citenamefont {Prakash}, \citenamefont
  {Chandra},\ and\ \citenamefont {Vachaspati}}]{prakash761448}%
  \BibitemOpen
  \bibfield  {author} {\bibinfo {author} {\bibfnamefont {H.}~\bibnamefont
  {Prakash}}, \bibinfo {author} {\bibfnamefont {N.}~\bibnamefont {Chandra}}, \
  and\ \bibinfo {author} {\bibnamefont {Vachaspati}},\ }\href@noop {}
  {\bibfield  {journal} {\bibinfo  {journal} {Indian J. Pure Appl. Phys.}\
  }\textbf {\bibinfo {volume} {14}},\ \bibinfo {pages} {48} (\bibinfo {year}
  {1976}{\natexlab{b}})}\BibitemShut {NoStop}%
\bibitem [{\citenamefont {Yuen}\ and\ \citenamefont
  {Shapiro}(1978)}]{yuen7824657}%
  \BibitemOpen
  \bibfield  {author} {\bibinfo {author} {\bibfnamefont {H.~P.}\ \bibnamefont
  {Yuen}}\ and\ \bibinfo {author} {\bibfnamefont {J.~H.}\ \bibnamefont
  {Shapiro}},\ }\href@noop {} {\bibfield  {journal} {\bibinfo  {journal} {IEEE
  Trans. Inf. Theory}\ }\textbf {\bibinfo {volume} {24}},\ \bibinfo {pages}
  {657} (\bibinfo {year} {1978})}\BibitemShut {NoStop}%
\bibitem [{\citenamefont {Shapiro}(1980)}]{shapiro805351}%
  \BibitemOpen
  \bibfield  {author} {\bibinfo {author} {\bibfnamefont {J.~H.}\ \bibnamefont
  {Shapiro}},\ }\href@noop {} {\bibfield  {journal} {\bibinfo  {journal} {Opt.
  Lett.}\ }\textbf {\bibinfo {volume} {5}},\ \bibinfo {pages} {351} (\bibinfo
  {year} {1980})}\BibitemShut {NoStop}%
\bibitem [{\citenamefont {Caves}(1981)}]{caves81231693}%
  \BibitemOpen
  \bibfield  {author} {\bibinfo {author} {\bibfnamefont {C.~M.}\ \bibnamefont
  {Caves}},\ }\href@noop {} {\bibfield  {journal} {\bibinfo  {journal} {Phys.
  Rev. D}\ }\textbf {\bibinfo {volume} {23}},\ \bibinfo {pages} {1693}
  (\bibinfo {year} {1981})}\BibitemShut {NoStop}%
\bibitem [{\citenamefont {Pace}\ \emph {et~al.}(1993)\citenamefont {Pace},
  \citenamefont {Collett},\ and\ \citenamefont {Walls}}]{pace93473173}%
  \BibitemOpen
  \bibfield  {author} {\bibinfo {author} {\bibfnamefont {A.~F.}\ \bibnamefont
  {Pace}}, \bibinfo {author} {\bibfnamefont {M.~J.}\ \bibnamefont {Collett}}, \
  and\ \bibinfo {author} {\bibfnamefont {D.~F.}\ \bibnamefont {Walls}},\
  }\href@noop {} {\bibfield  {journal} {\bibinfo  {journal} {Phys. Rev. A}\
  }\textbf {\bibinfo {volume} {47}},\ \bibinfo {pages} {3173} (\bibinfo {year}
  {1993})}\BibitemShut {NoStop}%
\bibitem [{\citenamefont {McKenzie}\ \emph {et~al.}(2004)\citenamefont
  {McKenzie}, \citenamefont {Grosse}, \citenamefont {Bowen}, \citenamefont
  {Whitcomb}, \citenamefont {Gray}, \citenamefont {McClelland},\ and\
  \citenamefont {Lam}}]{mckenzie0493161105}%
  \BibitemOpen
  \bibfield  {author} {\bibinfo {author} {\bibfnamefont {K.}~\bibnamefont
  {McKenzie}}, \bibinfo {author} {\bibfnamefont {N.}~\bibnamefont {Grosse}},
  \bibinfo {author} {\bibfnamefont {W.~P.}\ \bibnamefont {Bowen}}, \bibinfo
  {author} {\bibfnamefont {S.~E.}\ \bibnamefont {Whitcomb}}, \bibinfo {author}
  {\bibfnamefont {M.~B.}\ \bibnamefont {Gray}}, \bibinfo {author}
  {\bibfnamefont {D.~E.}\ \bibnamefont {McClelland}}, \ and\ \bibinfo {author}
  {\bibfnamefont {P.~K.}\ \bibnamefont {Lam}},\ }\href@noop {} {\bibfield
  {journal} {\bibinfo  {journal} {Phys. Rev. Lett.}\ }\textbf {\bibinfo
  {volume} {93}},\ \bibinfo {pages} {161105} (\bibinfo {year}
  {2004})}\BibitemShut {NoStop}%
\bibitem [{\citenamefont {Caves}\ and\ \citenamefont
  {Schumaker}(1985)}]{caves85313068}%
  \BibitemOpen
  \bibfield  {author} {\bibinfo {author} {\bibfnamefont {C.~M.}\ \bibnamefont
  {Caves}}\ and\ \bibinfo {author} {\bibfnamefont {B.~L.}\ \bibnamefont
  {Schumaker}},\ }\href@noop {} {\bibfield  {journal} {\bibinfo  {journal}
  {Phys. Rev. A}\ }\textbf {\bibinfo {volume} {31}},\ \bibinfo {pages} {3068}
  (\bibinfo {year} {1985})}\BibitemShut {NoStop}%
\bibitem [{\citenamefont {Schumaker}\ and\ \citenamefont
  {Caves}(1985)}]{schumaker85313093}%
  \BibitemOpen
  \bibfield  {author} {\bibinfo {author} {\bibfnamefont {B.~L.}\ \bibnamefont
  {Schumaker}}\ and\ \bibinfo {author} {\bibfnamefont {C.~M.}\ \bibnamefont
  {Caves}},\ }\href@noop {} {\bibfield  {journal} {\bibinfo  {journal} {Phys.
  Rev. A}\ }\textbf {\bibinfo {volume} {31}},\ \bibinfo {pages} {3093}
  (\bibinfo {year} {1985})}\BibitemShut {NoStop}%
\bibitem [{\citenamefont {Saleh}\ and\ \citenamefont
  {Teich}(1987)}]{saleh87582656}%
  \BibitemOpen
  \bibfield  {author} {\bibinfo {author} {\bibfnamefont {B.~E.~A.}\
  \bibnamefont {Saleh}}\ and\ \bibinfo {author} {\bibfnamefont {M.~C.}\
  \bibnamefont {Teich}},\ }\href@noop {} {\bibfield  {journal} {\bibinfo
  {journal} {Phys. Rev. Lett.}\ }\textbf {\bibinfo {volume} {58}},\ \bibinfo
  {pages} {2656} (\bibinfo {year} {1987})}\BibitemShut {NoStop}%
\bibitem [{\citenamefont {Braunstein}\ and\ \citenamefont
  {Kimble}(1998)}]{braunstein9880869}%
  \BibitemOpen
  \bibfield  {author} {\bibinfo {author} {\bibfnamefont {S.~L.}\ \bibnamefont
  {Braunstein}}\ and\ \bibinfo {author} {\bibfnamefont {H.~J.}\ \bibnamefont
  {Kimble}},\ }\href@noop {} {\bibfield  {journal} {\bibinfo  {journal} {Phys.
  Rev. Lett.}\ }\textbf {\bibinfo {volume} {80}},\ \bibinfo {pages} {869}
  (\bibinfo {year} {1998})}\BibitemShut {NoStop}%
\bibitem [{\citenamefont {Zhang}\ and\ \citenamefont
  {Peng}(2000)}]{zhang0062064302}%
  \BibitemOpen
  \bibfield  {author} {\bibinfo {author} {\bibfnamefont {J.}~\bibnamefont
  {Zhang}}\ and\ \bibinfo {author} {\bibfnamefont {K.~C.}\ \bibnamefont
  {Peng}},\ }\href@noop {} {\bibfield  {journal} {\bibinfo  {journal} {Phys.
  Rev. A}\ }\textbf {\bibinfo {volume} {62}},\ \bibinfo {pages} {064302}
  (\bibinfo {year} {2000})}\BibitemShut {NoStop}%
\bibitem [{\citenamefont {Bowen}\ \emph {et~al.}(2003)\citenamefont {Bowen},
  \citenamefont {Lam},\ and\ \citenamefont {Ralph}}]{bowen0350801}%
  \BibitemOpen
  \bibfield  {author} {\bibinfo {author} {\bibfnamefont {W.~P.}\ \bibnamefont
  {Bowen}}, \bibinfo {author} {\bibfnamefont {P.~K.}\ \bibnamefont {Lam}}, \
  and\ \bibinfo {author} {\bibfnamefont {T.~C.}\ \bibnamefont {Ralph}},\
  }\href@noop {} {\bibfield  {journal} {\bibinfo  {journal} {J. Mod. Opt.}\
  }\textbf {\bibinfo {volume} {50}},\ \bibinfo {pages} {801} (\bibinfo {year}
  {2003})}\BibitemShut {NoStop}%
\bibitem [{\citenamefont {Ban}(1999)}]{ban9919}%
  \BibitemOpen
  \bibfield  {author} {\bibinfo {author} {\bibfnamefont {M.}~\bibnamefont
  {Ban}},\ }\href@noop {} {\bibfield  {journal} {\bibinfo  {journal} {J. Opt.
  B: Quantum Semiclass. Opt.}\ }\textbf {\bibinfo {volume} {1}},\ \bibinfo
  {pages} {L9} (\bibinfo {year} {1999})}\BibitemShut {NoStop}%
\bibitem [{\citenamefont {Braunstein}\ and\ \citenamefont
  {Kimble}(2000)}]{braunstein0061042302}%
  \BibitemOpen
  \bibfield  {author} {\bibinfo {author} {\bibfnamefont {S.~L.}\ \bibnamefont
  {Braunstein}}\ and\ \bibinfo {author} {\bibfnamefont {H.~J.}\ \bibnamefont
  {Kimble}},\ }\href@noop {} {\bibfield  {journal} {\bibinfo  {journal} {Phys.
  Rev. A}\ }\textbf {\bibinfo {volume} {61}},\ \bibinfo {pages} {042302}
  (\bibinfo {year} {2000})}\BibitemShut {NoStop}%
\bibitem [{\citenamefont {Hillery}(2000)}]{hillery0061022309}%
  \BibitemOpen
  \bibfield  {author} {\bibinfo {author} {\bibfnamefont {M.}~\bibnamefont
  {Hillery}},\ }\href@noop {} {\bibfield  {journal} {\bibinfo  {journal} {Phys.
  Rev. A}\ }\textbf {\bibinfo {volume} {61}},\ \bibinfo {pages} {022309}
  (\bibinfo {year} {2000})}\BibitemShut {NoStop}%
\bibitem [{\citenamefont {Treps}\ \emph {et~al.}(2004)\citenamefont {Treps},
  \citenamefont {Grosse}, \citenamefont {Bowen}, \citenamefont {Hsu},
  \citenamefont {Ma\^{i}tre}, \citenamefont {Fabre}, \citenamefont {Bachor},\
  and\ \citenamefont {Lam}}]{treps046664}%
  \BibitemOpen
  \bibfield  {author} {\bibinfo {author} {\bibfnamefont {N.}~\bibnamefont
  {Treps}}, \bibinfo {author} {\bibfnamefont {N.}~\bibnamefont {Grosse}},
  \bibinfo {author} {\bibfnamefont {W.~P.}\ \bibnamefont {Bowen}}, \bibinfo
  {author} {\bibfnamefont {M.~T.~L.}\ \bibnamefont {Hsu}}, \bibinfo {author}
  {\bibfnamefont {A.}~\bibnamefont {Ma\^{i}tre}}, \bibinfo {author}
  {\bibfnamefont {C.}~\bibnamefont {Fabre}}, \bibinfo {author} {\bibfnamefont
  {H.-A.}\ \bibnamefont {Bachor}}, \ and\ \bibinfo {author} {\bibfnamefont
  {P.~K.}\ \bibnamefont {Lam}},\ }\href@noop {} {\bibfield  {journal} {\bibinfo
   {journal} {J. Opt. B: Quantum Semiclass.}\ }\textbf {\bibinfo {volume}
  {6}},\ \bibinfo {pages} {S664} (\bibinfo {year} {2004})}\BibitemShut
  {NoStop}%
\bibitem [{\citenamefont {Hsu}\ \emph {et~al.}(2006)\citenamefont {Hsu},
  \citenamefont {Delaubert}, \citenamefont {Bowen}, \citenamefont {Fabre},
  \citenamefont {Bachor},\ and\ \citenamefont {Lam}}]{hsu06421001}%
  \BibitemOpen
  \bibfield  {author} {\bibinfo {author} {\bibfnamefont {M.~T.~L.}\
  \bibnamefont {Hsu}}, \bibinfo {author} {\bibfnamefont {V.}~\bibnamefont
  {Delaubert}}, \bibinfo {author} {\bibfnamefont {W.~P.}\ \bibnamefont
  {Bowen}}, \bibinfo {author} {\bibfnamefont {C.}~\bibnamefont {Fabre}},
  \bibinfo {author} {\bibfnamefont {H.-A.}\ \bibnamefont {Bachor}}, \ and\
  \bibinfo {author} {\bibfnamefont {P.~K.}\ \bibnamefont {Lam}},\ }\href@noop
  {} {\bibfield  {journal} {\bibinfo  {journal} {IEEE J. Quantum Electron.}\
  }\textbf {\bibinfo {volume} {42}},\ \bibinfo {pages} {1001} (\bibinfo {year}
  {2006})}\BibitemShut {NoStop}%
\bibitem [{\citenamefont {Kozlovskii}(2007)}]{kozlovskii073774}%
  \BibitemOpen
  \bibfield  {author} {\bibinfo {author} {\bibfnamefont {A.~V.}\ \bibnamefont
  {Kozlovskii}},\ }\href@noop {} {\bibfield  {journal} {\bibinfo  {journal}
  {Quantum Electronics}\ }\textbf {\bibinfo {volume} {37}},\ \bibinfo {pages}
  {74} (\bibinfo {year} {2007})}\BibitemShut {NoStop}%
\bibitem [{\citenamefont {Laflamme}\ and\ \citenamefont
  {Clerk}(2011)}]{laflamme1183033803}%
  \BibitemOpen
  \bibfield  {author} {\bibinfo {author} {\bibfnamefont {C.}~\bibnamefont
  {Laflamme}}\ and\ \bibinfo {author} {\bibfnamefont {A.~A.}\ \bibnamefont
  {Clerk}},\ }\href@noop {} {\bibfield  {journal} {\bibinfo  {journal} {Phys.
  Rev. A}\ }\textbf {\bibinfo {volume} {83}},\ \bibinfo {pages} {033803}
  (\bibinfo {year} {2011})}\BibitemShut {NoStop}%
\bibitem [{\citenamefont {Simaan}\ and\ \citenamefont
  {Loudon}(1975)}]{simaan758539}%
  \BibitemOpen
  \bibfield  {author} {\bibinfo {author} {\bibfnamefont {H.~D.}\ \bibnamefont
  {Simaan}}\ and\ \bibinfo {author} {\bibfnamefont {R.}~\bibnamefont
  {Loudon}},\ }\href@noop {} {\bibfield  {journal} {\bibinfo  {journal} {J.
  Phys. A: Math. Gen.}\ }\textbf {\bibinfo {volume} {8}},\ \bibinfo {pages}
  {539} (\bibinfo {year} {1975})}\BibitemShut {NoStop}%
\bibitem [{\citenamefont {Loudon}(1984)}]{loudon844967}%
  \BibitemOpen
  \bibfield  {author} {\bibinfo {author} {\bibfnamefont {R.}~\bibnamefont
  {Loudon}},\ }\href@noop {} {\bibfield  {journal} {\bibinfo  {journal} {Opt.
  Commun.}\ }\textbf {\bibinfo {volume} {49}},\ \bibinfo {pages} {67} (\bibinfo
  {year} {1984})}\BibitemShut {NoStop}%
\bibitem [{\citenamefont {Milburn}\ and\ \citenamefont
  {Walls}(1981)}]{milburn8139401}%
  \BibitemOpen
  \bibfield  {author} {\bibinfo {author} {\bibfnamefont {G.~J.}\ \bibnamefont
  {Milburn}}\ and\ \bibinfo {author} {\bibfnamefont {D.~F.}\ \bibnamefont
  {Walls}},\ }\href@noop {} {\bibfield  {journal} {\bibinfo  {journal} {Opt.
  Commun.}\ }\textbf {\bibinfo {volume} {39}},\ \bibinfo {pages} {401}
  (\bibinfo {year} {1981})}\BibitemShut {NoStop}%
\bibitem [{\citenamefont {Becker}\ \emph {et~al.}(1982)\citenamefont {Becker},
  \citenamefont {Scully},\ and\ \citenamefont {Zubairy}}]{becker8248475}%
  \BibitemOpen
  \bibfield  {author} {\bibinfo {author} {\bibfnamefont {W.}~\bibnamefont
  {Becker}}, \bibinfo {author} {\bibfnamefont {M.~O.}\ \bibnamefont {Scully}},
  \ and\ \bibinfo {author} {\bibfnamefont {M.~S.}\ \bibnamefont {Zubairy}},\
  }\href@noop {} {\bibfield  {journal} {\bibinfo  {journal} {Phys. Rev. Lett.}\
  }\textbf {\bibinfo {volume} {48}},\ \bibinfo {pages} {475} (\bibinfo {year}
  {1982})}\BibitemShut {NoStop}%
\bibitem [{\citenamefont {Mandel}(1982{\natexlab{a}})}]{mandel8232437}%
  \BibitemOpen
  \bibfield  {author} {\bibinfo {author} {\bibfnamefont {L.}~\bibnamefont
  {Mandel}},\ }\href@noop {} {\bibfield  {journal} {\bibinfo  {journal} {Opt.
  Commun.}\ }\textbf {\bibinfo {volume} {42}},\ \bibinfo {pages} {437}
  (\bibinfo {year} {1982}{\natexlab{a}})}\BibitemShut {NoStop}%
\bibitem [{\citenamefont {Lugiato}\ \emph {et~al.}(1983)\citenamefont
  {Lugiato}, \citenamefont {Strini},\ and\ \citenamefont
  {Martini}}]{lugiato838256}%
  \BibitemOpen
  \bibfield  {author} {\bibinfo {author} {\bibfnamefont {L.~A.}\ \bibnamefont
  {Lugiato}}, \bibinfo {author} {\bibfnamefont {G.}~\bibnamefont {Strini}}, \
  and\ \bibinfo {author} {\bibfnamefont {F.~D.}\ \bibnamefont {Martini}},\
  }\href@noop {} {\bibfield  {journal} {\bibinfo  {journal} {Opt. Lett.}\
  }\textbf {\bibinfo {volume} {8}},\ \bibinfo {pages} {256} (\bibinfo {year}
  {1983})}\BibitemShut {NoStop}%
\bibitem [{\citenamefont {Milburn}\ and\ \citenamefont
  {Walls}(1983)}]{milburn8327392}%
  \BibitemOpen
  \bibfield  {author} {\bibinfo {author} {\bibfnamefont {G.~J.}\ \bibnamefont
  {Milburn}}\ and\ \bibinfo {author} {\bibfnamefont {D.~F.}\ \bibnamefont
  {Walls}},\ }\href@noop {} {\bibfield  {journal} {\bibinfo  {journal} {Phys.
  Rev. A}\ }\textbf {\bibinfo {volume} {27}},\ \bibinfo {pages} {392} (\bibinfo
  {year} {1983})}\BibitemShut {NoStop}%
\bibitem [{\citenamefont {Bondurant}\ \emph {et~al.}(1984)\citenamefont
  {Bondurant}, \citenamefont {Kumar}, \citenamefont {Shapiro},\ and\
  \citenamefont {Maeda}}]{bondurant8430343}%
  \BibitemOpen
  \bibfield  {author} {\bibinfo {author} {\bibfnamefont {R.~S.}\ \bibnamefont
  {Bondurant}}, \bibinfo {author} {\bibfnamefont {P.}~\bibnamefont {Kumar}},
  \bibinfo {author} {\bibfnamefont {J.~H.}\ \bibnamefont {Shapiro}}, \ and\
  \bibinfo {author} {\bibfnamefont {M.}~\bibnamefont {Maeda}},\ }\href@noop {}
  {\bibfield  {journal} {\bibinfo  {journal} {Phys. Rev. A}\ }\textbf {\bibinfo
  {volume} {30}},\ \bibinfo {pages} {343} (\bibinfo {year} {1984})}\BibitemShut
  {NoStop}%
\bibitem [{\citenamefont {Reid}\ and\ \citenamefont
  {Walls}(1985)}]{reid85311622}%
  \BibitemOpen
  \bibfield  {author} {\bibinfo {author} {\bibfnamefont {M.~D.}\ \bibnamefont
  {Reid}}\ and\ \bibinfo {author} {\bibfnamefont {D.~F.}\ \bibnamefont
  {Walls}},\ }\href@noop {} {\bibfield  {journal} {\bibinfo  {journal} {Phys.
  Rev. A}\ }\textbf {\bibinfo {volume} {31}},\ \bibinfo {pages} {1622}
  (\bibinfo {year} {1985})}\BibitemShut {NoStop}%
\bibitem [{\citenamefont {Perinova}\ and\ \citenamefont
  {Tiebel}(1984)}]{perinova8450401}%
  \BibitemOpen
  \bibfield  {author} {\bibinfo {author} {\bibfnamefont {V.}~\bibnamefont
  {Perinova}}\ and\ \bibinfo {author} {\bibfnamefont {R.}~\bibnamefont
  {Tiebel}},\ }\href@noop {} {\bibfield  {journal} {\bibinfo  {journal} {Opt.
  Commun.}\ }\textbf {\bibinfo {volume} {50}},\ \bibinfo {pages} {401}
  (\bibinfo {year} {1984})}\BibitemShut {NoStop}%
\bibitem [{\citenamefont {Ficek}\ \emph {et~al.}(1984)\citenamefont {Ficek},
  \citenamefont {Tanas},\ and\ \citenamefont {Kielich}}]{ficek84292004}%
  \BibitemOpen
  \bibfield  {author} {\bibinfo {author} {\bibfnamefont {Z.}~\bibnamefont
  {Ficek}}, \bibinfo {author} {\bibfnamefont {R.}~\bibnamefont {Tanas}}, \ and\
  \bibinfo {author} {\bibfnamefont {S.}~\bibnamefont {Kielich}},\ }\href@noop
  {} {\bibfield  {journal} {\bibinfo  {journal} {Phys. Rev. A}\ }\textbf
  {\bibinfo {volume} {29}},\ \bibinfo {pages} {2004} (\bibinfo {year}
  {1984})}\BibitemShut {NoStop}%
\bibitem [{\citenamefont {Li}\ and\ \citenamefont {Wu}(2001)}]{li0119797}%
  \BibitemOpen
  \bibfield  {author} {\bibinfo {author} {\bibfnamefont {H.}~\bibnamefont
  {Li}}\ and\ \bibinfo {author} {\bibfnamefont {L.}~\bibnamefont {Wu}},\
  }\href@noop {} {\bibfield  {journal} {\bibinfo  {journal} {Opt. Commun.}\
  }\textbf {\bibinfo {volume} {197}},\ \bibinfo {pages} {97} (\bibinfo {year}
  {2001})}\BibitemShut {NoStop}%
\bibitem [{\citenamefont {Zheng}(2007)}]{zheng07273460}%
  \BibitemOpen
  \bibfield  {author} {\bibinfo {author} {\bibfnamefont {S.-B.}\ \bibnamefont
  {Zheng}},\ }\href@noop {} {\bibfield  {journal} {\bibinfo  {journal} {Opt.
  Commun.}\ }\textbf {\bibinfo {volume} {273}},\ \bibinfo {pages} {460}
  (\bibinfo {year} {2007})}\BibitemShut {NoStop}%
\bibitem [{\citenamefont {Prakash}\ and\ \citenamefont
  {Kumar}(2003{\natexlab{a}})}]{prakash03319305}%
  \BibitemOpen
  \bibfield  {author} {\bibinfo {author} {\bibfnamefont {H.}~\bibnamefont
  {Prakash}}\ and\ \bibinfo {author} {\bibfnamefont {P.}~\bibnamefont
  {Kumar}},\ }\href@noop {} {\bibfield  {journal} {\bibinfo  {journal} {Physica
  A}\ }\textbf {\bibinfo {volume} {319}},\ \bibinfo {pages} {305} (\bibinfo
  {year} {2003}{\natexlab{a}})}\BibitemShut {NoStop}%
\bibitem [{\citenamefont {Prakash}\ and\ \citenamefont
  {Kumar}(2004)}]{prakash04341201}%
  \BibitemOpen
  \bibfield  {author} {\bibinfo {author} {\bibfnamefont {H.}~\bibnamefont
  {Prakash}}\ and\ \bibinfo {author} {\bibfnamefont {P.}~\bibnamefont
  {Kumar}},\ }\href@noop {} {\bibfield  {journal} {\bibinfo  {journal} {Physica
  A}\ }\textbf {\bibinfo {volume} {341}},\ \bibinfo {pages} {201} (\bibinfo
  {year} {2004})}\BibitemShut {NoStop}%
\bibitem [{\citenamefont {Prakash}\ and\ \citenamefont
  {Kumar}(2011)}]{prakash111221058}%
  \BibitemOpen
  \bibfield  {author} {\bibinfo {author} {\bibfnamefont {H.}~\bibnamefont
  {Prakash}}\ and\ \bibinfo {author} {\bibfnamefont {P.}~\bibnamefont
  {Kumar}},\ }\href@noop {} {\bibfield  {journal} {\bibinfo  {journal} {Optik}\
  }\textbf {\bibinfo {volume} {122}},\ \bibinfo {pages} {1058} (\bibinfo {year}
  {2011})}\BibitemShut {NoStop}%
\bibitem [{\citenamefont {Prakash}\ and\ \citenamefont
  {Mishra}(2010)}]{prakash10352212}%
  \BibitemOpen
  \bibfield  {author} {\bibinfo {author} {\bibfnamefont {H.}~\bibnamefont
  {Prakash}}\ and\ \bibinfo {author} {\bibfnamefont {D.~K.}\ \bibnamefont
  {Mishra}},\ }\href@noop {} {\bibfield  {journal} {\bibinfo  {journal} {Opt.
  Lett.}\ }\textbf {\bibinfo {volume} {35}},\ \bibinfo {pages} {2212} (\bibinfo
  {year} {2010})}\BibitemShut {NoStop}%
\bibitem [{\citenamefont {Carruthers}\ and\ \citenamefont
  {Nieto}(1965)}]{carruthers6514387}%
  \BibitemOpen
  \bibfield  {author} {\bibinfo {author} {\bibfnamefont {P.}~\bibnamefont
  {Carruthers}}\ and\ \bibinfo {author} {\bibfnamefont {M.~M.}\ \bibnamefont
  {Nieto}},\ }\href@noop {} {\bibfield  {journal} {\bibinfo  {journal} {Phys.
  Rev. Lett.}\ }\textbf {\bibinfo {volume} {14}},\ \bibinfo {pages} {387}
  (\bibinfo {year} {1965})}\BibitemShut {NoStop}%
\bibitem [{\citenamefont {Mandel}\ and\ \citenamefont
  {Wolf}(1995)}]{mandelandwolf95}%
  \BibitemOpen
  \bibfield  {author} {\bibinfo {author} {\bibfnamefont {L.}~\bibnamefont
  {Mandel}}\ and\ \bibinfo {author} {\bibfnamefont {E.}~\bibnamefont {Wolf}},\
  }\href@noop {} {\emph {\bibinfo {title} {Optical Coherence and Quantum
  Optics}}}\ (\bibinfo  {publisher} {Cambridge University Press},\ \bibinfo
  {year} {1995})\BibitemShut {NoStop}%
\bibitem [{\citenamefont {Scully}\ and\ \citenamefont
  {Zubairy}(1997)}]{scullyandzubairy97}%
  \BibitemOpen
  \bibfield  {author} {\bibinfo {author} {\bibfnamefont {M.~O.}\ \bibnamefont
  {Scully}}\ and\ \bibinfo {author} {\bibfnamefont {M.~S.}\ \bibnamefont
  {Zubairy}},\ }\href@noop {} {\emph {\bibinfo {title} {Quantum Optics}}}\
  (\bibinfo  {publisher} {Cambridge University Press},\ \bibinfo {year}
  {1997})\BibitemShut {NoStop}%
\bibitem [{\citenamefont {Gerry}\ and\ \citenamefont
  {Knight}(2005)}]{gerryandknight05}%
  \BibitemOpen
  \bibfield  {author} {\bibinfo {author} {\bibfnamefont {C.~C.}\ \bibnamefont
  {Gerry}}\ and\ \bibinfo {author} {\bibfnamefont {P.~L.}\ \bibnamefont
  {Knight}},\ }\href@noop {} {\emph {\bibinfo {title} {Introductory Quantum
  Optics}}}\ (\bibinfo  {publisher} {Cambridge University Press},\ \bibinfo
  {year} {2005})\BibitemShut {NoStop}%
\bibitem [{\citenamefont {Walls}\ and\ \citenamefont
  {Milburn}(2008)}]{wallsandmilburn08}%
  \BibitemOpen
  \bibfield  {author} {\bibinfo {author} {\bibfnamefont {D.~F.}\ \bibnamefont
  {Walls}}\ and\ \bibinfo {author} {\bibfnamefont {G.~J.}\ \bibnamefont
  {Milburn}},\ }\href@noop {} {\emph {\bibinfo {title} {Quantum Optics}}},\
  \bibinfo {edition} {2nd}\ ed.\ (\bibinfo  {publisher} {Springer},\ \bibinfo
  {year} {2008})\BibitemShut {NoStop}%
\bibitem [{asl()}]{asl}%
  \BibitemOpen
  \href {http://www.rp-photonics.com/amplitude_squeezed_light.html} {\enquote
  {\bibinfo {title} {Amplitude squeezed light},}\ }\BibitemShut {NoStop}%
\bibitem [{ssl()}]{ssl}%
  \BibitemOpen
  \href {http://www.rp-photonics.com/squeezed_states_of_light.html} {\enquote
  {\bibinfo {title} {Squeezed states of light},}\ }\BibitemShut {NoStop}%
\bibitem [{\citenamefont {Hong}\ and\ \citenamefont
  {Mandel}(1985{\natexlab{a}})}]{hong8554323}%
  \BibitemOpen
  \bibfield  {author} {\bibinfo {author} {\bibfnamefont {C.~K.}\ \bibnamefont
  {Hong}}\ and\ \bibinfo {author} {\bibfnamefont {L.}~\bibnamefont {Mandel}},\
  }\href@noop {} {\bibfield  {journal} {\bibinfo  {journal} {Phys. Rev. Lett.}\
  }\textbf {\bibinfo {volume} {54}},\ \bibinfo {pages} {323} (\bibinfo {year}
  {1985}{\natexlab{a}})}\BibitemShut {NoStop}%
\bibitem [{\citenamefont {Hong}\ and\ \citenamefont
  {Mandel}(1985{\natexlab{b}})}]{hong8532974}%
  \BibitemOpen
  \bibfield  {author} {\bibinfo {author} {\bibfnamefont {C.~K.}\ \bibnamefont
  {Hong}}\ and\ \bibinfo {author} {\bibfnamefont {L.}~\bibnamefont {Mandel}},\
  }\href@noop {} {\bibfield  {journal} {\bibinfo  {journal} {Phys. Rev. A}\
  }\textbf {\bibinfo {volume} {32}},\ \bibinfo {pages} {974} (\bibinfo {year}
  {1985}{\natexlab{b}})}\BibitemShut {NoStop}%
\bibitem [{\citenamefont {Garc\'{i}a-Fern\'{a}ndez}\ \emph
  {et~al.}(1986)\citenamefont {Garc\'{i}a-Fern\'{a}ndez}, \citenamefont
  {{\uppercase{d}e Los}~Terreros}, \citenamefont {Bermejo},\ and\ \citenamefont
  {Santoro}}]{fernandez86118400}%
  \BibitemOpen
  \bibfield  {author} {\bibinfo {author} {\bibfnamefont {P.}~\bibnamefont
  {Garc\'{i}a-Fern\'{a}ndez}}, \bibinfo {author} {\bibfnamefont {L.~S.}\
  \bibnamefont {{\uppercase{d}e Los}~Terreros}}, \bibinfo {author}
  {\bibfnamefont {F.~J.}\ \bibnamefont {Bermejo}}, \ and\ \bibinfo {author}
  {\bibfnamefont {J.}~\bibnamefont {Santoro}},\ }\href@noop {} {\bibfield
  {journal} {\bibinfo  {journal} {Phys. Lett. A}\ }\textbf {\bibinfo {volume}
  {118}},\ \bibinfo {pages} {400} (\bibinfo {year} {1986})}\BibitemShut
  {NoStop}%
\bibitem [{\citenamefont {Tombesi}\ and\ \citenamefont
  {Mecozzi}(1988)}]{tombesi88374778}%
  \BibitemOpen
  \bibfield  {author} {\bibinfo {author} {\bibfnamefont {P.}~\bibnamefont
  {Tombesi}}\ and\ \bibinfo {author} {\bibfnamefont {A.}~\bibnamefont
  {Mecozzi}},\ }\href@noop {} {\bibfield  {journal} {\bibinfo  {journal} {Phys.
  Rev. A}\ }\textbf {\bibinfo {volume} {37}},\ \bibinfo {pages} {4778}
  (\bibinfo {year} {1988})}\BibitemShut {NoStop}%
\bibitem [{\citenamefont {Marian}(1991)}]{marian91443325}%
  \BibitemOpen
  \bibfield  {author} {\bibinfo {author} {\bibfnamefont {P.}~\bibnamefont
  {Marian}},\ }\href@noop {} {\bibfield  {journal} {\bibinfo  {journal} {Phys.
  Rev. A}\ }\textbf {\bibinfo {volume} {44}},\ \bibinfo {pages} {3325}
  (\bibinfo {year} {1991})}\BibitemShut {NoStop}%
\bibitem [{\citenamefont {Marian}(1992)}]{marian92452044}%
  \BibitemOpen
  \bibfield  {author} {\bibinfo {author} {\bibfnamefont {P.}~\bibnamefont
  {Marian}},\ }\href@noop {} {\bibfield  {journal} {\bibinfo  {journal} {Phys.
  Rev. A}\ }\textbf {\bibinfo {volume} {45}},\ \bibinfo {pages} {2044}
  (\bibinfo {year} {1992})}\BibitemShut {NoStop}%
\bibitem [{\citenamefont {Wang}\ \emph {et~al.}(2000)\citenamefont {Wang},
  \citenamefont {Liu},\ and\ \citenamefont {Zhan}}]{wang00392583}%
  \BibitemOpen
  \bibfield  {author} {\bibinfo {author} {\bibfnamefont {J.-S.}\ \bibnamefont
  {Wang}}, \bibinfo {author} {\bibfnamefont {T.-K.}\ \bibnamefont {Liu}}, \
  and\ \bibinfo {author} {\bibfnamefont {M.-S.}\ \bibnamefont {Zhan}},\
  }\href@noop {} {\bibfield  {journal} {\bibinfo  {journal} {Int. J. Theor.
  Phys.}\ }\textbf {\bibinfo {volume} {39}},\ \bibinfo {pages} {2583} (\bibinfo
  {year} {2000})}\BibitemShut {NoStop}%
\bibitem [{\citenamefont {Prakash}\ and\ \citenamefont
  {Kumar}(2003{\natexlab{b}})}]{praksah03342769}%
  \BibitemOpen
  \bibfield  {author} {\bibinfo {author} {\bibfnamefont {H.}~\bibnamefont
  {Prakash}}\ and\ \bibinfo {author} {\bibfnamefont {P.}~\bibnamefont
  {Kumar}},\ }\href@noop {} {\bibfield  {journal} {\bibinfo  {journal} {Acta
  Physica Polonica B}\ }\textbf {\bibinfo {volume} {34}},\ \bibinfo {pages}
  {2769} (\bibinfo {year} {2003}{\natexlab{b}})}\BibitemShut {NoStop}%
\bibitem [{\citenamefont {Mishra}(2010)}]{mishra102833284}%
  \BibitemOpen
  \bibfield  {author} {\bibinfo {author} {\bibfnamefont {D.~K.}\ \bibnamefont
  {Mishra}},\ }\href@noop {} {\bibfield  {journal} {\bibinfo  {journal} {Opt.
  Commun.}\ }\textbf {\bibinfo {volume} {283}},\ \bibinfo {pages} {3284}
  (\bibinfo {year} {2010})}\BibitemShut {NoStop}%
\bibitem [{\citenamefont {Prakash}\ \emph {et~al.}(2011)\citenamefont
  {Prakash}, \citenamefont {Kumar},\ and\ \citenamefont
  {Kumar}}]{prakash11284289}%
  \BibitemOpen
  \bibfield  {author} {\bibinfo {author} {\bibfnamefont {H.}~\bibnamefont
  {Prakash}}, \bibinfo {author} {\bibfnamefont {R.}~\bibnamefont {Kumar}}, \
  and\ \bibinfo {author} {\bibfnamefont {P.}~\bibnamefont {Kumar}},\
  }\href@noop {} {\bibfield  {journal} {\bibinfo  {journal} {Opt. Commun.}\
  }\textbf {\bibinfo {volume} {284}},\ \bibinfo {pages} {289} (\bibinfo {year}
  {2011})}\BibitemShut {NoStop}%
\bibitem [{\citenamefont {Hillery}(1987{\natexlab{a}})}]{hillery8762135}%
  \BibitemOpen
  \bibfield  {author} {\bibinfo {author} {\bibfnamefont {M.}~\bibnamefont
  {Hillery}},\ }\href@noop {} {\bibfield  {journal} {\bibinfo  {journal} {Opt.
  Commun.}\ }\textbf {\bibinfo {volume} {62}},\ \bibinfo {pages} {135}
  (\bibinfo {year} {1987}{\natexlab{a}})}\BibitemShut {NoStop}%
\bibitem [{\citenamefont {Hillery}(1987{\natexlab{b}})}]{hillery87363796}%
  \BibitemOpen
  \bibfield  {author} {\bibinfo {author} {\bibfnamefont {M.}~\bibnamefont
  {Hillery}},\ }\href@noop {} {\bibfield  {journal} {\bibinfo  {journal} {Phys.
  Rev. A}\ }\textbf {\bibinfo {volume} {36}},\ \bibinfo {pages} {3796}
  (\bibinfo {year} {1987}{\natexlab{b}})}\BibitemShut {NoStop}%
\bibitem [{\citenamefont {Zhang}\ \emph {et~al.}(1990)\citenamefont {Zhang},
  \citenamefont {Xu}, \citenamefont {Chai},\ and\ \citenamefont
  {Li}}]{zhang9015027}%
  \BibitemOpen
  \bibfield  {author} {\bibinfo {author} {\bibfnamefont {Z.-M.}\ \bibnamefont
  {Zhang}}, \bibinfo {author} {\bibfnamefont {L.}~\bibnamefont {Xu}}, \bibinfo
  {author} {\bibfnamefont {J.-L.}\ \bibnamefont {Chai}}, \ and\ \bibinfo
  {author} {\bibfnamefont {F.-L.}\ \bibnamefont {Li}},\ }\href@noop {}
  {\bibfield  {journal} {\bibinfo  {journal} {Phys. Lett. A}\ }\textbf
  {\bibinfo {volume} {150}},\ \bibinfo {pages} {27} (\bibinfo {year}
  {1990})}\BibitemShut {NoStop}%
\bibitem [{\citenamefont {Zhan}(1991)}]{zhan91160498}%
  \BibitemOpen
  \bibfield  {author} {\bibinfo {author} {\bibfnamefont {Y.}~\bibnamefont
  {Zhan}},\ }\href@noop {} {\bibfield  {journal} {\bibinfo  {journal} {Phys.
  Lett. A}\ }\textbf {\bibinfo {volume} {160}},\ \bibinfo {pages} {498}
  (\bibinfo {year} {1991})}\BibitemShut {NoStop}%
\bibitem [{\citenamefont {Zhan}(1992)}]{zhan9246686}%
  \BibitemOpen
  \bibfield  {author} {\bibinfo {author} {\bibfnamefont {Y.}~\bibnamefont
  {Zhan}},\ }\href@noop {} {\bibfield  {journal} {\bibinfo  {journal} {Phys.
  Rev. A}\ }\textbf {\bibinfo {volume} {46}},\ \bibinfo {pages} {686} (\bibinfo
  {year} {1992})}\BibitemShut {NoStop}%
\bibitem [{\citenamefont {Gerry}\ and\ \citenamefont
  {Vrscay}(1988)}]{gerry88371779}%
  \BibitemOpen
  \bibfield  {author} {\bibinfo {author} {\bibfnamefont {C.~C.}\ \bibnamefont
  {Gerry}}\ and\ \bibinfo {author} {\bibfnamefont {E.~R.}\ \bibnamefont
  {Vrscay}},\ }\href@noop {} {\bibfield  {journal} {\bibinfo  {journal} {Phys.
  Rev. A}\ }\textbf {\bibinfo {volume} {37}},\ \bibinfo {pages} {1779}
  (\bibinfo {year} {1988})}\BibitemShut {NoStop}%
\bibitem [{\citenamefont {Mahran}\ and\ \citenamefont
  {Obada}(1989)}]{mahran89404476}%
  \BibitemOpen
  \bibfield  {author} {\bibinfo {author} {\bibfnamefont {M.~H.}\ \bibnamefont
  {Mahran}}\ and\ \bibinfo {author} {\bibfnamefont {A.-S.~F.}\ \bibnamefont
  {Obada}},\ }\href@noop {} {\bibfield  {journal} {\bibinfo  {journal} {Phys.
  Rev. A}\ }\textbf {\bibinfo {volume} {40}},\ \bibinfo {pages} {4476}
  (\bibinfo {year} {1989})}\BibitemShut {NoStop}%
\bibitem [{\citenamefont {Obada}\ and\ \citenamefont
  {Abdel-Hafez}(1992)}]{obada929199}%
  \BibitemOpen
  \bibfield  {author} {\bibinfo {author} {\bibfnamefont {A.-S.~F.}\
  \bibnamefont {Obada}}\ and\ \bibinfo {author} {\bibfnamefont {A.~M.}\
  \bibnamefont {Abdel-Hafez}},\ }\href@noop {} {\bibfield  {journal} {\bibinfo
  {journal} {Opt. Commun.}\ }\textbf {\bibinfo {volume} {91}},\ \bibinfo
  {pages} {99} (\bibinfo {year} {1992})}\BibitemShut {NoStop}%
\bibitem [{\citenamefont {Prakash}\ and\ \citenamefont
  {Kumar}(2007{\natexlab{a}})}]{prakash213621}%
  \BibitemOpen
  \bibfield  {author} {\bibinfo {author} {\bibfnamefont {H.}~\bibnamefont
  {Prakash}}\ and\ \bibinfo {author} {\bibfnamefont {R.}~\bibnamefont
  {Kumar}},\ }\href@noop {} {\bibfield  {journal} {\bibinfo  {journal} {Int. J.
  Mod. Phys. B}\ }\textbf {\bibinfo {volume} {21}},\ \bibinfo {pages} {3621}
  (\bibinfo {year} {2007}{\natexlab{a}})}\BibitemShut {NoStop}%
\bibitem [{\citenamefont {Giri}\ and\ \citenamefont {Gupta}(2004)}]{giri04691}%
  \BibitemOpen
  \bibfield  {author} {\bibinfo {author} {\bibfnamefont {D.~K.}\ \bibnamefont
  {Giri}}\ and\ \bibinfo {author} {\bibfnamefont {P.~S.}\ \bibnamefont
  {Gupta}},\ }\href@noop {} {\bibfield  {journal} {\bibinfo  {journal} {J. Opt.
  B: Quantum Semiclass. Opt.}\ }\textbf {\bibinfo {volume} {6}},\ \bibinfo
  {pages} {91} (\bibinfo {year} {2004})}\BibitemShut {NoStop}%
\bibitem [{\citenamefont {Giri}\ and\ \citenamefont
  {Gupta}(2008)}]{giri0822219}%
  \BibitemOpen
  \bibfield  {author} {\bibinfo {author} {\bibfnamefont {D.~K.}\ \bibnamefont
  {Giri}}\ and\ \bibinfo {author} {\bibfnamefont {P.~S.}\ \bibnamefont
  {Gupta}},\ }\href@noop {} {\bibfield  {journal} {\bibinfo  {journal} {Mod.
  Phys. Lett. B}\ }\textbf {\bibinfo {volume} {22}},\ \bibinfo {pages} {219}
  (\bibinfo {year} {2008})}\BibitemShut {NoStop}%
\bibitem [{\citenamefont {Prakash}\ and\ \citenamefont
  {Kumar}(2006)}]{prakash06201458}%
  \BibitemOpen
  \bibfield  {author} {\bibinfo {author} {\bibfnamefont {H.}~\bibnamefont
  {Prakash}}\ and\ \bibinfo {author} {\bibfnamefont {P.}~\bibnamefont
  {Kumar}},\ }\href@noop {} {\bibfield  {journal} {\bibinfo  {journal} {Int. J.
  Mod. Phys. B}\ }\textbf {\bibinfo {volume} {20}},\ \bibinfo {pages} {1458}
  (\bibinfo {year} {2006})}\BibitemShut {NoStop}%
\bibitem [{\citenamefont {Wu}\ \emph {et~al.}(2007)\citenamefont {Wu},
  \citenamefont {Cheng}, \citenamefont {Zhang},\ and\ \citenamefont
  {Cheng}}]{wu0747933}%
  \BibitemOpen
  \bibfield  {author} {\bibinfo {author} {\bibfnamefont {Z.-X.}\ \bibnamefont
  {Wu}}, \bibinfo {author} {\bibfnamefont {Z.}~\bibnamefont {Cheng}}, \bibinfo
  {author} {\bibfnamefont {Y.-M.}\ \bibnamefont {Zhang}}, \ and\ \bibinfo
  {author} {\bibfnamefont {Z.-Z.}\ \bibnamefont {Cheng}},\ }\href@noop {}
  {\bibfield  {journal} {\bibinfo  {journal} {Chin.Theor. Phys.}\ }\textbf
  {\bibinfo {volume} {47}},\ \bibinfo {pages} {933} (\bibinfo {year}
  {2007})}\BibitemShut {NoStop}%
\bibitem [{\citenamefont {Marian}(1997)}]{marian97553051}%
  \BibitemOpen
  \bibfield  {author} {\bibinfo {author} {\bibfnamefont {P.}~\bibnamefont
  {Marian}},\ }\href@noop {} {\bibfield  {journal} {\bibinfo  {journal} {Phys.
  Rev. A}\ }\textbf {\bibinfo {volume} {55}},\ \bibinfo {pages} {3051}
  (\bibinfo {year} {1997})}\BibitemShut {NoStop}%
\bibitem [{\citenamefont {Prakash}\ and\ \citenamefont
  {Kumar}(2008)}]{prakash0846359}%
  \BibitemOpen
  \bibfield  {author} {\bibinfo {author} {\bibfnamefont {H.}~\bibnamefont
  {Prakash}}\ and\ \bibinfo {author} {\bibfnamefont {P.}~\bibnamefont
  {Kumar}},\ }\href@noop {} {\bibfield  {journal} {\bibinfo  {journal} {Euro.
  Phys. J. D}\ }\textbf {\bibinfo {volume} {46}},\ \bibinfo {pages} {359}
  (\bibinfo {year} {2008})}\BibitemShut {NoStop}%
\bibitem [{\citenamefont {Prakash}\ and\ \citenamefont
  {Mishra}(2005)}]{prakash0538665}%
  \BibitemOpen
  \bibfield  {author} {\bibinfo {author} {\bibfnamefont {H.}~\bibnamefont
  {Prakash}}\ and\ \bibinfo {author} {\bibfnamefont {D.~K.}\ \bibnamefont
  {Mishra}},\ }\href@noop {} {\bibfield  {journal} {\bibinfo  {journal} {J.
  Phys. B: At. Mol. Opt. Phys.}\ }\textbf {\bibinfo {volume} {38}},\ \bibinfo
  {pages} {665} (\bibinfo {year} {2005})}\BibitemShut {NoStop}%
\bibitem [{\citenamefont {Mishra}(2007)}]{mishra07112859}%
  \BibitemOpen
  \bibfield  {author} {\bibinfo {author} {\bibfnamefont {D.~K.}\ \bibnamefont
  {Mishra}},\ }\href@noop {} {\bibfield  {journal} {\bibinfo  {journal} {Acta
  Physica Polonica A}\ }\textbf {\bibinfo {volume} {112}},\ \bibinfo {pages}
  {859} (\bibinfo {year} {2007})}\BibitemShut {NoStop}%
\bibitem [{\citenamefont {Kumar}\ and\ \citenamefont
  {Gupta}(1996)}]{kumar9681053}%
  \BibitemOpen
  \bibfield  {author} {\bibinfo {author} {\bibfnamefont {A.}~\bibnamefont
  {Kumar}}\ and\ \bibinfo {author} {\bibfnamefont {P.~S.}\ \bibnamefont
  {Gupta}},\ }\href@noop {} {\bibfield  {journal} {\bibinfo  {journal} {Quantum
  Semiclass. Opt.}\ }\textbf {\bibinfo {volume} {8}},\ \bibinfo {pages} {1053}
  (\bibinfo {year} {1996})}\BibitemShut {NoStop}%
\bibitem [{\citenamefont {Giri}\ and\ \citenamefont
  {Gupta}(2006)}]{giri06202265}%
  \BibitemOpen
  \bibfield  {author} {\bibinfo {author} {\bibfnamefont {D.~K.}\ \bibnamefont
  {Giri}}\ and\ \bibinfo {author} {\bibfnamefont {P.~S.}\ \bibnamefont
  {Gupta}},\ }\href@noop {} {\bibfield  {journal} {\bibinfo  {journal} {Int. J.
  Mod. Phys. B}\ }\textbf {\bibinfo {volume} {20}},\ \bibinfo {pages} {2265}
  (\bibinfo {year} {2006})}\BibitemShut {NoStop}%
\bibitem [{\citenamefont {Rani}\ \emph
  {et~al.}(2007{\natexlab{a}})\citenamefont {Rani}, \citenamefont {Lal},\ and\
  \citenamefont {Singh}}]{rani0739157}%
  \BibitemOpen
  \bibfield  {author} {\bibinfo {author} {\bibfnamefont {S.}~\bibnamefont
  {Rani}}, \bibinfo {author} {\bibfnamefont {J.}~\bibnamefont {Lal}}, \ and\
  \bibinfo {author} {\bibfnamefont {N.}~\bibnamefont {Singh}},\ }\href@noop {}
  {\bibfield  {journal} {\bibinfo  {journal} {Opt. Quant. Electron}\ }\textbf
  {\bibinfo {volume} {39}},\ \bibinfo {pages} {157} (\bibinfo {year}
  {2007}{\natexlab{a}})}\BibitemShut {NoStop}%
\bibitem [{\citenamefont {Sen}\ and\ \citenamefont
  {Mandal}(2008)}]{sen08551697}%
  \BibitemOpen
  \bibfield  {author} {\bibinfo {author} {\bibfnamefont {B.}~\bibnamefont
  {Sen}}\ and\ \bibinfo {author} {\bibfnamefont {S.}~\bibnamefont {Mandal}},\
  }\href@noop {} {\bibfield  {journal} {\bibinfo  {journal} {J. Mod. Opt.}\
  }\textbf {\bibinfo {volume} {55}},\ \bibinfo {pages} {1697} (\bibinfo {year}
  {2008})}\BibitemShut {NoStop}%
\bibitem [{\citenamefont {Du}\ and\ \citenamefont {Gong}(1993)}]{du93482198}%
  \BibitemOpen
  \bibfield  {author} {\bibinfo {author} {\bibfnamefont {S.}~\bibnamefont
  {Du}}\ and\ \bibinfo {author} {\bibfnamefont {C.}~\bibnamefont {Gong}},\
  }\href@noop {} {\bibfield  {journal} {\bibinfo  {journal} {Phys. Rev. A}\
  }\textbf {\bibinfo {volume} {48}},\ \bibinfo {pages} {2198} (\bibinfo {year}
  {1993})}\BibitemShut {NoStop}%
\bibitem [{\citenamefont {Wang}\ \emph
  {et~al.}(1995{\natexlab{a}})\citenamefont {Wang}, \citenamefont {Sui},\ and\
  \citenamefont {Wang}}]{wang957917}%
  \BibitemOpen
  \bibfield  {author} {\bibinfo {author} {\bibfnamefont {J.}~\bibnamefont
  {Wang}}, \bibinfo {author} {\bibfnamefont {Q.}~\bibnamefont {Sui}}, \ and\
  \bibinfo {author} {\bibfnamefont {C.}~\bibnamefont {Wang}},\ }\href@noop {}
  {\bibfield  {journal} {\bibinfo  {journal} {Quant. Semi. Opt.}\ }\textbf
  {\bibinfo {volume} {7}},\ \bibinfo {pages} {917} (\bibinfo {year}
  {1995}{\natexlab{a}})}\BibitemShut {NoStop}%
\bibitem [{\citenamefont {Rani}\ \emph
  {et~al.}(2007{\natexlab{b}})\citenamefont {Rani}, \citenamefont {Lal},\ and\
  \citenamefont {Singh}}]{rani0739735}%
  \BibitemOpen
  \bibfield  {author} {\bibinfo {author} {\bibfnamefont {S.}~\bibnamefont
  {Rani}}, \bibinfo {author} {\bibfnamefont {J.}~\bibnamefont {Lal}}, \ and\
  \bibinfo {author} {\bibfnamefont {N.}~\bibnamefont {Singh}},\ }\href@noop {}
  {\bibfield  {journal} {\bibinfo  {journal} {Opt. Quant. Electron}\ }\textbf
  {\bibinfo {volume} {39}},\ \bibinfo {pages} {735} (\bibinfo {year}
  {2007}{\natexlab{b}})}\BibitemShut {NoStop}%
\bibitem [{\citenamefont {Rani}\ \emph {et~al.}(2008)\citenamefont {Rani},
  \citenamefont {Lal},\ and\ \citenamefont {Singh}}]{rani08281341}%
  \BibitemOpen
  \bibfield  {author} {\bibinfo {author} {\bibfnamefont {S.}~\bibnamefont
  {Rani}}, \bibinfo {author} {\bibfnamefont {J.}~\bibnamefont {Lal}}, \ and\
  \bibinfo {author} {\bibfnamefont {N.}~\bibnamefont {Singh}},\ }\href@noop {}
  {\bibfield  {journal} {\bibinfo  {journal} {Opt. Commun.}\ }\textbf {\bibinfo
  {volume} {281}},\ \bibinfo {pages} {341} (\bibinfo {year}
  {2008})}\BibitemShut {NoStop}%
\bibitem [{\citenamefont {Rani}\ \emph {et~al.}(2009)\citenamefont {Rani},
  \citenamefont {Lal},\ and\ \citenamefont {Singh}}]{rani09232681}%
  \BibitemOpen
  \bibfield  {author} {\bibinfo {author} {\bibfnamefont {S.}~\bibnamefont
  {Rani}}, \bibinfo {author} {\bibfnamefont {J.}~\bibnamefont {Lal}}, \ and\
  \bibinfo {author} {\bibfnamefont {N.}~\bibnamefont {Singh}},\ }\href@noop {}
  {\bibfield  {journal} {\bibinfo  {journal} {Mod. Phys. Lett. B}\ }\textbf
  {\bibinfo {volume} {23}},\ \bibinfo {pages} {2681} (\bibinfo {year}
  {2009})}\BibitemShut {NoStop}%
\bibitem [{\citenamefont {Giri}\ and\ \citenamefont
  {Gupta}(2005)}]{giri05191943}%
  \BibitemOpen
  \bibfield  {author} {\bibinfo {author} {\bibfnamefont {D.~K.}\ \bibnamefont
  {Giri}}\ and\ \bibinfo {author} {\bibfnamefont {P.~S.}\ \bibnamefont
  {Gupta}},\ }\href@noop {} {\bibfield  {journal} {\bibinfo  {journal} {Int. J.
  Mod. Phys. B}\ }\textbf {\bibinfo {volume} {19}},\ \bibinfo {pages} {1943}
  (\bibinfo {year} {2005})}\BibitemShut {NoStop}%
\bibitem [{\citenamefont {Rani}\ \emph
  {et~al.}(2007{\natexlab{c}})\citenamefont {Rani}, \citenamefont {Lal},\ and\
  \citenamefont {Singh}}]{rani07277427}%
  \BibitemOpen
  \bibfield  {author} {\bibinfo {author} {\bibfnamefont {S.}~\bibnamefont
  {Rani}}, \bibinfo {author} {\bibfnamefont {J.}~\bibnamefont {Lal}}, \ and\
  \bibinfo {author} {\bibfnamefont {N.}~\bibnamefont {Singh}},\ }\href@noop {}
  {\bibfield  {journal} {\bibinfo  {journal} {Opt. Commun.}\ }\textbf {\bibinfo
  {volume} {277}},\ \bibinfo {pages} {427} (\bibinfo {year}
  {2007}{\natexlab{c}})}\BibitemShut {NoStop}%
\bibitem [{\citenamefont {Wang}\ \emph
  {et~al.}(1995{\natexlab{b}})\citenamefont {Wang}, \citenamefont {Wang},
  \citenamefont {Sun},\ and\ \citenamefont {He}}]{wang954247}%
  \BibitemOpen
  \bibfield  {author} {\bibinfo {author} {\bibfnamefont {J.-S.}\ \bibnamefont
  {Wang}}, \bibinfo {author} {\bibfnamefont {C.-K.}\ \bibnamefont {Wang}},
  \bibinfo {author} {\bibfnamefont {J.-Z.}\ \bibnamefont {Sun}}, \ and\
  \bibinfo {author} {\bibfnamefont {J.-Y.}\ \bibnamefont {He}},\ }\href@noop {}
  {\bibfield  {journal} {\bibinfo  {journal} {Acta Physica Sinica}\ }\textbf
  {\bibinfo {volume} {4}},\ \bibinfo {pages} {247} (\bibinfo {year}
  {1995}{\natexlab{b}})}\BibitemShut {NoStop}%
\bibitem [{\citenamefont {Lynch}(1986)}]{lynch86334431}%
  \BibitemOpen
  \bibfield  {author} {\bibinfo {author} {\bibfnamefont {R.}~\bibnamefont
  {Lynch}},\ }\href@noop {} {\bibfield  {journal} {\bibinfo  {journal} {Phys.
  Rev. A}\ }\textbf {\bibinfo {volume} {33}},\ \bibinfo {pages} {4431}
  (\bibinfo {year} {1986})}\BibitemShut {NoStop}%
\bibitem [{\citenamefont {Hong}\ and\ \citenamefont
  {Mandel}(1986)}]{hong86334432}%
  \BibitemOpen
  \bibfield  {author} {\bibinfo {author} {\bibfnamefont {C.~K.}\ \bibnamefont
  {Hong}}\ and\ \bibinfo {author} {\bibfnamefont {L.}~\bibnamefont {Mandel}},\
  }\href@noop {} {\bibfield  {journal} {\bibinfo  {journal} {Phys. Rev. A}\
  }\textbf {\bibinfo {volume} {33}},\ \bibinfo {pages} {4432} (\bibinfo {year}
  {1986})}\BibitemShut {NoStop}%
\bibitem [{\citenamefont {Lynch}(1994)}]{lynch94492800}%
  \BibitemOpen
  \bibfield  {author} {\bibinfo {author} {\bibfnamefont {R.}~\bibnamefont
  {Lynch}},\ }\href@noop {} {\bibfield  {journal} {\bibinfo  {journal} {Phys.
  Rev. A}\ }\textbf {\bibinfo {volume} {49}},\ \bibinfo {pages} {2800}
  (\bibinfo {year} {1994})}\BibitemShut {NoStop}%
\bibitem [{\citenamefont {Bužek}\ and\ \citenamefont
  {Jex}(1990)}]{buzek90414079}%
  \BibitemOpen
  \bibfield  {author} {\bibinfo {author} {\bibfnamefont {V.}~\bibnamefont
  {Bužek}}\ and\ \bibinfo {author} {\bibfnamefont {I.}~\bibnamefont {Jex}},\
  }\href@noop {} {\bibfield  {journal} {\bibinfo  {journal} {Phys. Rev. A}\
  }\textbf {\bibinfo {volume} {41}},\ \bibinfo {pages} {4079} (\bibinfo {year}
  {1990})}\BibitemShut {NoStop}%
\bibitem [{\citenamefont {Hillery}(1989)}]{hillery89403147}%
  \BibitemOpen
  \bibfield  {author} {\bibinfo {author} {\bibfnamefont {M.}~\bibnamefont
  {Hillery}},\ }\href@noop {} {\bibfield  {journal} {\bibinfo  {journal} {Phys.
  Rev. A}\ }\textbf {\bibinfo {volume} {40}},\ \bibinfo {pages} {3147}
  (\bibinfo {year} {1989})}\BibitemShut {NoStop}%
\bibitem [{\citenamefont {Kumar}\ and\ \citenamefont
  {Gupta}(1997)}]{kumar97136441}%
  \BibitemOpen
  \bibfield  {author} {\bibinfo {author} {\bibfnamefont {A.}~\bibnamefont
  {Kumar}}\ and\ \bibinfo {author} {\bibfnamefont {P.~S.}\ \bibnamefont
  {Gupta}},\ }\href@noop {} {\bibfield  {journal} {\bibinfo  {journal} {Opt.
  Commun.}\ }\textbf {\bibinfo {volume} {136}},\ \bibinfo {pages} {441}
  (\bibinfo {year} {1997})}\BibitemShut {NoStop}%
\bibitem [{\citenamefont {Kumar}\ and\ \citenamefont
  {Gupta}(1998)}]{kumar9810485}%
  \BibitemOpen
  \bibfield  {author} {\bibinfo {author} {\bibfnamefont {A.}~\bibnamefont
  {Kumar}}\ and\ \bibinfo {author} {\bibfnamefont {P.~S.}\ \bibnamefont
  {Gupta}},\ }\href@noop {} {\bibfield  {journal} {\bibinfo  {journal} {Quantum
  Semiclass. Opt.}\ }\textbf {\bibinfo {volume} {10}},\ \bibinfo {pages} {485}
  (\bibinfo {year} {1998})}\BibitemShut {NoStop}%
\bibitem [{\citenamefont {Prakash}\ and\ \citenamefont
  {Mishra}(2007{\natexlab{a}})}]{prakash0745363}%
  \BibitemOpen
  \bibfield  {author} {\bibinfo {author} {\bibfnamefont {H.}~\bibnamefont
  {Prakash}}\ and\ \bibinfo {author} {\bibfnamefont {D.~K.}\ \bibnamefont
  {Mishra}},\ }\href@noop {} {\bibfield  {journal} {\bibinfo  {journal} {Eur.
  Phys. J. D}\ }\textbf {\bibinfo {volume} {45}},\ \bibinfo {pages} {363}
  (\bibinfo {year} {2007}{\natexlab{a}})}\BibitemShut {NoStop}%
\bibitem [{\citenamefont {Prakash}\ and\ \citenamefont
  {Kumar}(2005{\natexlab{a}})}]{prakash057757}%
  \BibitemOpen
  \bibfield  {author} {\bibinfo {author} {\bibfnamefont {H.}~\bibnamefont
  {Prakash}}\ and\ \bibinfo {author} {\bibfnamefont {R.}~\bibnamefont
  {Kumar}},\ }\href@noop {} {\bibfield  {journal} {\bibinfo  {journal} {J. Opt.
  B: Qunatum Semiclass. Opt.}\ }\textbf {\bibinfo {volume} {7}},\ \bibinfo
  {pages} {S757} (\bibinfo {year} {2005}{\natexlab{a}})}\BibitemShut {NoStop}%
\bibitem [{\citenamefont {Prakash}\ and\ \citenamefont
  {Kumar}(2007{\natexlab{b}})}]{prakash0742475}%
  \BibitemOpen
  \bibfield  {author} {\bibinfo {author} {\bibfnamefont {H.}~\bibnamefont
  {Prakash}}\ and\ \bibinfo {author} {\bibfnamefont {R.}~\bibnamefont
  {Kumar}},\ }\href@noop {} {\bibfield  {journal} {\bibinfo  {journal} {Eur.
  Phys. J. D}\ }\textbf {\bibinfo {volume} {42}},\ \bibinfo {pages} {475}
  (\bibinfo {year} {2007}{\natexlab{b}})}\BibitemShut {NoStop}%
\bibitem [{\citenamefont {Chirkin}\ \emph {et~al.}(1993)\citenamefont
  {Chirkin}, \citenamefont {Orlov},\ and\ \citenamefont
  {Parashchuk}}]{chirkin9323870}%
  \BibitemOpen
  \bibfield  {author} {\bibinfo {author} {\bibfnamefont {A.~S.}\ \bibnamefont
  {Chirkin}}, \bibinfo {author} {\bibfnamefont {A.~A.}\ \bibnamefont {Orlov}},
  \ and\ \bibinfo {author} {\bibfnamefont {D.~Y.}\ \bibnamefont {Parashchuk}},\
  }\href@noop {} {\bibfield  {journal} {\bibinfo  {journal} {Quantum
  Electron.}\ }\textbf {\bibinfo {volume} {23}},\ \bibinfo {pages} {870}
  (\bibinfo {year} {1993})}\BibitemShut {NoStop}%
\bibitem [{\citenamefont {Prakash}\ and\ \citenamefont
  {Shukla}(2009)}]{psicop2009}%
  \BibitemOpen
  \bibfield  {author} {\bibinfo {author} {\bibfnamefont {R.}~\bibnamefont
  {Prakash}}\ and\ \bibinfo {author} {\bibfnamefont {N.}~\bibnamefont
  {Shukla}},\ }in\ \href@noop {} {\emph {\bibinfo {booktitle} {Proceedings of
  International Conference on Optics and Photonics}}}\ (\bibinfo {address}
  {CSIO, Chandigarh, India},\ \bibinfo {year} {2009})\ p.\ \bibinfo {pages}
  {244}\BibitemShut {NoStop}%
\bibitem [{\citenamefont {Prakash}\ and\ \citenamefont
  {Shukla}(2011)}]{prakash112843568}%
  \BibitemOpen
  \bibfield  {author} {\bibinfo {author} {\bibfnamefont {R.}~\bibnamefont
  {Prakash}}\ and\ \bibinfo {author} {\bibfnamefont {N.}~\bibnamefont
  {Shukla}},\ }\href@noop {} {\bibfield  {journal} {\bibinfo  {journal} {Opt.
  Commun.}\ }\textbf {\bibinfo {volume} {284}},\ \bibinfo {pages} {3568}
  (\bibinfo {year} {2011})}\BibitemShut {NoStop}%
\bibitem [{\citenamefont {Mandel}(1982{\natexlab{b}})}]{mandel8249136}%
  \BibitemOpen
  \bibfield  {author} {\bibinfo {author} {\bibfnamefont {L.}~\bibnamefont
  {Mandel}},\ }\href@noop {} {\bibfield  {journal} {\bibinfo  {journal} {Phys.
  Rev. Lett.}\ }\textbf {\bibinfo {volume} {49}},\ \bibinfo {pages} {136}
  (\bibinfo {year} {1982}{\natexlab{b}})}\BibitemShut {NoStop}%
\bibitem [{\citenamefont {Mandel}(1979)}]{mandel794205}%
  \BibitemOpen
  \bibfield  {author} {\bibinfo {author} {\bibfnamefont {L.}~\bibnamefont
  {Mandel}},\ }\href@noop {} {\bibfield  {journal} {\bibinfo  {journal} {Opt.
  Lett.}\ }\textbf {\bibinfo {volume} {4}},\ \bibinfo {pages} {205} (\bibinfo
  {year} {1979})}\BibitemShut {NoStop}%
\bibitem [{\citenamefont {Short}\ and\ \citenamefont
  {Mandel}(1983)}]{short8351384}%
  \BibitemOpen
  \bibfield  {author} {\bibinfo {author} {\bibfnamefont {R.}~\bibnamefont
  {Short}}\ and\ \bibinfo {author} {\bibfnamefont {L.}~\bibnamefont {Mandel}},\
  }\href@noop {} {\bibfield  {journal} {\bibinfo  {journal} {Phys. Rev. Lett.}\
  }\textbf {\bibinfo {volume} {51}},\ \bibinfo {pages} {384} (\bibinfo {year}
  {1983})}\BibitemShut {NoStop}%
\bibitem [{\citenamefont {Kumar}\ and\ \citenamefont
  {Prakash}(2010)}]{kumar1088181}%
  \BibitemOpen
  \bibfield  {author} {\bibinfo {author} {\bibfnamefont {R.}~\bibnamefont
  {Kumar}}\ and\ \bibinfo {author} {\bibfnamefont {H.}~\bibnamefont
  {Prakash}},\ }\href@noop {} {\bibfield  {journal} {\bibinfo  {journal} {Can.
  J. Phys.}\ }\textbf {\bibinfo {volume} {88}},\ \bibinfo {pages} {181}
  (\bibinfo {year} {2010})}\BibitemShut {NoStop}%
\bibitem [{\citenamefont {Prakash}\ and\ \citenamefont
  {Kumar}(2005{\natexlab{b}})}]{prakash057786}%
  \BibitemOpen
  \bibfield  {author} {\bibinfo {author} {\bibfnamefont {H.}~\bibnamefont
  {Prakash}}\ and\ \bibinfo {author} {\bibfnamefont {P.}~\bibnamefont
  {Kumar}},\ }\href@noop {} {\bibfield  {journal} {\bibinfo  {journal} {J. Opt.
  B: Quantum Semiclass. Opt.}\ }\textbf {\bibinfo {volume} {7}},\ \bibinfo
  {pages} {S786} (\bibinfo {year} {2005}{\natexlab{b}})}\BibitemShut {NoStop}%
\bibitem [{\citenamefont {Prakash}\ and\ \citenamefont
  {Mishra}(2006)}]{prakash06392291}%
  \BibitemOpen
  \bibfield  {author} {\bibinfo {author} {\bibfnamefont {H.}~\bibnamefont
  {Prakash}}\ and\ \bibinfo {author} {\bibfnamefont {D.~K.}\ \bibnamefont
  {Mishra}},\ }\href@noop {} {\bibfield  {journal} {\bibinfo  {journal} {J.
  Phys. B: At. Mol. Opt. Phys.}\ }\textbf {\bibinfo {volume} {39}},\ \bibinfo
  {pages} {2291} (\bibinfo {year} {2006})}\BibitemShut {NoStop}%
\bibitem [{\citenamefont {Prakash}\ and\ \citenamefont
  {Mishra}(2007{\natexlab{b}})}]{prakash07402531}%
  \BibitemOpen
  \bibfield  {author} {\bibinfo {author} {\bibfnamefont {H.}~\bibnamefont
  {Prakash}}\ and\ \bibinfo {author} {\bibfnamefont {D.~K.}\ \bibnamefont
  {Mishra}},\ }\href@noop {} {\bibfield  {journal} {\bibinfo  {journal} {J.
  Phys. B: At. Mol. Opt. Phys.}\ }\textbf {\bibinfo {volume} {40}},\ \bibinfo
  {pages} {2531} (\bibinfo {year} {2007}{\natexlab{b}})}\BibitemShut {NoStop}%
\bibitem [{\citenamefont {Prakash}\ \emph {et~al.}(2010)\citenamefont
  {Prakash}, \citenamefont {Kumar},\ and\ \citenamefont
  {Mishra}}]{prakash10245547}%
  \BibitemOpen
  \bibfield  {author} {\bibinfo {author} {\bibfnamefont {H.}~\bibnamefont
  {Prakash}}, \bibinfo {author} {\bibfnamefont {P.}~\bibnamefont {Kumar}}, \
  and\ \bibinfo {author} {\bibfnamefont {D.~K.}\ \bibnamefont {Mishra}},\
  }\href@noop {} {\bibfield  {journal} {\bibinfo  {journal} {Int. J. Mod. Phys.
  B}\ }\textbf {\bibinfo {volume} {24}},\ \bibinfo {pages} {5547} (\bibinfo
  {year} {2010})}\BibitemShut {NoStop}%
\bibitem [{\citenamefont {Shchukin}\ and\ \citenamefont
  {Vogel}(2005)}]{shchukin0572043808}%
  \BibitemOpen
  \bibfield  {author} {\bibinfo {author} {\bibfnamefont {E.~V.}\ \bibnamefont
  {Shchukin}}\ and\ \bibinfo {author} {\bibfnamefont {W.}~\bibnamefont
  {Vogel}},\ }\href@noop {} {\bibfield  {journal} {\bibinfo  {journal} {Phys.
  Rev. A}\ }\textbf {\bibinfo {volume} {72}},\ \bibinfo {pages} {043808}
  (\bibinfo {year} {2005})}\BibitemShut {NoStop}%
\bibitem [{\citenamefont {Vogel}\ and\ \citenamefont
  {Shchukin}(2007)}]{vogel0784012020}%
  \BibitemOpen
  \bibfield  {author} {\bibinfo {author} {\bibfnamefont {W.}~\bibnamefont
  {Vogel}}\ and\ \bibinfo {author} {\bibfnamefont {E.}~\bibnamefont
  {Shchukin}},\ }\href@noop {} {\bibfield  {journal} {\bibinfo  {journal} {J.
  Phys.: Conf. Ser.}\ }\textbf {\bibinfo {volume} {84}},\ \bibinfo {pages}
  {012020} (\bibinfo {year} {2007})}\BibitemShut {NoStop}%
\bibitem [{\citenamefont {Shchukin}\ and\ \citenamefont
  {Vogel}(2006)}]{shchukin0696200403}%
  \BibitemOpen
  \bibfield  {author} {\bibinfo {author} {\bibfnamefont {E.}~\bibnamefont
  {Shchukin}}\ and\ \bibinfo {author} {\bibfnamefont {W.}~\bibnamefont
  {Vogel}},\ }\href@noop {} {\bibfield  {journal} {\bibinfo  {journal} {Phys.
  Rev. Lett.}\ }\textbf {\bibinfo {volume} {96}},\ \bibinfo {pages} {200403}
  (\bibinfo {year} {2006})}\BibitemShut {NoStop}%
\bibitem [{\citenamefont {Penin}\ and\ \citenamefont
  {Sergienko}(1991)}]{penin91303582}%
  \BibitemOpen
  \bibfield  {author} {\bibinfo {author} {\bibfnamefont {A.~N.}\ \bibnamefont
  {Penin}}\ and\ \bibinfo {author} {\bibfnamefont {A.~V.}\ \bibnamefont
  {Sergienko}},\ }\href@noop {} {\bibfield  {journal} {\bibinfo  {journal}
  {Appl. Opt.}\ }\textbf {\bibinfo {volume} {30}},\ \bibinfo {pages} {3582}
  (\bibinfo {year} {1991})}\BibitemShut {NoStop}%
\bibitem [{\citenamefont {Castelletto}\ \emph {et~al.}(1995)\citenamefont
  {Castelletto}, \citenamefont {Godone}, \citenamefont {Novero},\ and\
  \citenamefont {Rastello}}]{castelletto9532501}%
  \BibitemOpen
  \bibfield  {author} {\bibinfo {author} {\bibfnamefont {S.}~\bibnamefont
  {Castelletto}}, \bibinfo {author} {\bibfnamefont {A.}~\bibnamefont {Godone}},
  \bibinfo {author} {\bibfnamefont {C.}~\bibnamefont {Novero}}, \ and\ \bibinfo
  {author} {\bibfnamefont {M.~L.}\ \bibnamefont {Rastello}},\ }\href@noop {}
  {\bibfield  {journal} {\bibinfo  {journal} {Metrologia}\ }\textbf {\bibinfo
  {volume} {32}},\ \bibinfo {pages} {501} (\bibinfo {year} {1995})}\BibitemShut
  {NoStop}%
\bibitem [{\citenamefont {Brida}\ \emph {et~al.}(1998)\citenamefont {Brida},
  \citenamefont {Castelletto}, \citenamefont {Novero},\ and\ \citenamefont
  {Rastello}}]{brida9835397}%
  \BibitemOpen
  \bibfield  {author} {\bibinfo {author} {\bibfnamefont {G.}~\bibnamefont
  {Brida}}, \bibinfo {author} {\bibfnamefont {S.}~\bibnamefont {Castelletto}},
  \bibinfo {author} {\bibfnamefont {C.}~\bibnamefont {Novero}}, \ and\ \bibinfo
  {author} {\bibfnamefont {M.~L.}\ \bibnamefont {Rastello}},\ }\href@noop {}
  {\bibfield  {journal} {\bibinfo  {journal} {Metrologia}\ }\textbf {\bibinfo
  {volume} {35}},\ \bibinfo {pages} {397} (\bibinfo {year} {1998})}\BibitemShut
  {NoStop}%
\bibitem [{\citenamefont {Migdall}\ \emph {et~al.}(2002)\citenamefont
  {Migdall}, \citenamefont {Castelletto}, \citenamefont {Degiovanni},\ and\
  \citenamefont {Rastello}}]{migdall02412914}%
  \BibitemOpen
  \bibfield  {author} {\bibinfo {author} {\bibfnamefont {A.}~\bibnamefont
  {Migdall}}, \bibinfo {author} {\bibfnamefont {S.}~\bibnamefont
  {Castelletto}}, \bibinfo {author} {\bibfnamefont {I.~P.}\ \bibnamefont
  {Degiovanni}}, \ and\ \bibinfo {author} {\bibfnamefont {M.~L.}\ \bibnamefont
  {Rastello}},\ }\href@noop {} {\bibfield  {journal} {\bibinfo  {journal}
  {Appl. Opt.}\ }\textbf {\bibinfo {volume} {41}},\ \bibinfo {pages} {2914}
  (\bibinfo {year} {2002})}\BibitemShut {NoStop}%
\bibitem [{\citenamefont {Lu}\ \emph {et~al.}(2002)\citenamefont {Lu},
  \citenamefont {Liu}, \citenamefont {Sun}, \citenamefont {Xu},\ and\
  \citenamefont {Jiang}}]{lu0213186}%
  \BibitemOpen
  \bibfield  {author} {\bibinfo {author} {\bibfnamefont {S.}~\bibnamefont
  {Lu}}, \bibinfo {author} {\bibfnamefont {B.}~\bibnamefont {Liu}}, \bibinfo
  {author} {\bibfnamefont {B.}~\bibnamefont {Sun}}, \bibinfo {author}
  {\bibfnamefont {Z.}~\bibnamefont {Xu}}, \ and\ \bibinfo {author}
  {\bibfnamefont {D.}~\bibnamefont {Jiang}},\ }\href@noop {} {\bibfield
  {journal} {\bibinfo  {journal} {Meas. Sci. Technol.}\ }\textbf {\bibinfo
  {volume} {13}},\ \bibinfo {pages} {186} (\bibinfo {year} {2002})}\BibitemShut
  {NoStop}%
\end{thebibliography}%

\end{document}